\documentclass[10pt]{article}
\pdfoutput=1 

\usepackage{jheppub} 

\usepackage[T1]{fontenc} 
\usepackage{enumitem}

\usepackage{xcolor}
\usepackage{xspace}
\usepackage{amssymb,amsthm,cancel,hyperref,graphicx,xcolor}
\usepackage{mathrsfs,wasysym}
\usepackage{booktabs}
\usepackage{subcaption}
\usepackage{tikz}
\usetikzlibrary{calc}
\usepackage{tabu}
\usepackage{pdflscape}
\usepackage{afterpage}
\usepackage{graphicx}
\usepackage{makecell}

\usepackage{todonotes}
\presetkeys{todonotes}{inline}{}
\usepackage{amsmath}
\usepackage{amssymb}
\usepackage{commath}
\usepackage{array}
\usepackage{calc}
\usepackage{longtable}
\usepackage{multirow}
\usepackage{physics}
\usepackage{tensor}
\usepackage{upgreek}
\usepackage{graphicx}
\graphicspath{{figures/}}
\usepackage{xspace}
\usepackage{listings}
\usepackage[section]{placeins}
\usepackage{
  pgf,
  tikz}
\usetikzlibrary{
  patterns,
  shapes.multipart,
  arrows,
  trees,
  scopes,
  decorations.pathreplacing,
  decorations.pathmorphing,
  decorations.markings,
  decorations.text,
  positioning,
  calc
}

\usepackage[utf8]{inputenc}
\usepackage{mciteplus}

\newcommand{\Sherpa}{S\protect\scalebox{0.8}{HERPA}\xspace}
\newcommand{\Rivet}{R\protect\scalebox{0.8}{IVET}\xspace}
\newcommand{\Herwig}{H\protect\scalebox{0.8}{ERWIG}\xspace}

\newcommand{\Pythia}{P\protect\scalebox{0.8}{YTHIA}\xspace}
\newcommand{\Amegic}{A\protect\scalebox{0.8}{MEGIC}\xspace}
\newcommand{\Comix}{C\protect\scalebox{0.8}{OMIX}\xspace}
\newcommand{\Collier}{C\protect\scalebox{0.8}{OLLIER}\xspace}

\newcommand{\MCatNLO}{MC\protect\scalebox{0.8}{@}NLO\xspace}

\newcommand{\OpenLoops}{\textsc{OpenLoops}}

\newcommand{\LEP}{L\protect\scalebox{0.8}{EP}\xspace}

\usepackage{color}
\definecolor{darkblue}{rgb}{0,0,0.5}
\definecolor{darkred}{rgb}{0.5,0,0}
\definecolor{darkgreen}{rgb}{0,0.5,0}

\preprint{\newline FERMILAB-PUB-26-0425-T\\IPPP/26/53\\MCNET-26-16}

\title{Reweighting Underlying Event and Colour Reconnection parameter variations in \protect\Sherpa}

\author[a]{Moritz Pabst,}
\author[b]{Max Knobbe,} 
\author[c]{Frank Krauss,}
\author[a]{Steffen Schumann}

\affiliation[a]{Institut f{\"u}r Theoretische Physik, Georg-August-Universit{\"a}t G\"ottingen,\\ Friedrich-Hund-Platz 1, 37077 G\"ottingen, Germany}
\affiliation[b]{Theoretical Physics Division, Fermi National Accelerator Laboratory, Batavia, IL 60510, USA}
\affiliation[c]{Institute for Particle Physics Phenomenology, Durham University, Durham DH1 3LE, UK}
\emailAdd{moritz.pabst@stud.uni-goettingen.de}
\emailAdd{mknobbe@fnal.gov}
\emailAdd{frank.krauss@durham.ac.uk}
\emailAdd{steffen.schumann@phys.uni-goettingen.de}

\abstract{
  We propose and validate a new method to trace the impact of parameter variations in the simulation
  of multi-parton interactions and colour reconnections in the \protect\Sherpa event generator.
  They are reflected, at an event-by-event basis, through relative weights with respect to the
  central production parameters that give rise to the generated events and distributions.
  Our method facilitates the \emph{tuning} of the Monte Carlo event generator at a dramatically
  reduced computational cost, alleviates parameter sensitivity studies, and enables robust quantification
  of parametric uncertainties \emph{on-the-fly}, one of the missing ingredients for future
  simulations of high-energy particle collisions. The method can easily be adapted to and implemented
  in other event generators. To illustrate its potential, we here consider combined tunes of the
  multi-parton-interaction and colour-reconnection models in \protect\Sherpa using LHC proton--proton
  collision data at $\sqrt{s}=7\,\text{TeV}$. We furthermore calibrate the energy-scaling behaviour
  of dimensionful model parameters based on $\sqrt{s}=13\,\text{TeV}$ LHC data and Tevatron data
  taken at $\sqrt{s}=1.96\,\text{TeV}$.
}

\begin{document}
\maketitle

\section{Introduction}

Detailed particle-level simulations that encapsulate theoretical expectations are a central ingredient to the analysis and interpretation of data at
high-energy collider experiments.
The widely used general-purpose Monte Carlo event generators (MCEGs) \Herwig~\cite{Bewick:2023tfi,Bellm:2025pcw},
\Pythia~\cite{Sjostrand:2007gs,Sjostrand:2014zea,Bierlich:2022pfr}, and
\Sherpa~\cite{Gleisberg:2008ta,Bothmann:2019yzt,Sherpa:2024mfk} provide detailed and complete particle-level predictions, accounting for the complex
evolution from partonic hard scatterings to hadronic final states by combining first-principles perturbative QCD calculations with phenomenological models
for non-perturbative (NP) effects~\cite{Buckley:2011ms,Campbell:2022qmc}.

The robust quantification of inherent theoretical uncertainties in the underlying physics models is crucial for precision phenomenology and the correct
interpretation of data, including, for example, evidence for new phenomena or constraints on them.
In the perturbative regime, on-the-fly event reweighting techniques facilitate the efficient evaluation of uncertainties in both the hard scattering
matrix elements and the subsequent parton shower, as they arise from variations in the strong coupling, scale choices, and parton-density functions
(PDFs)~\cite{Bellm:2016voq,Mrenna:2016sih,Bothmann:2016nao}.
These techniques have become indispensable for the estimation of perturbative uncertainties and have eliminated the computationally
costly need to regenerate events with varying settings.

However, many phenomenologically important uncertainties originate from the NP modelling of the underlying event, hadronisation, and (soft) colour
reconnections. These models depend on empirical parameters that must be calibrated, ``tuned'', to existing collider data.  This is typically achieved
through global fits using tools such as {\sc Professor}~\cite{Buckley:2009bj} and {\sc Apprentice}~\cite{Krishnamoorthy:2021nwv} or, more recently,
the technique of History Matching (HM) using Bayes Linear Emulators~\cite{Iskauskas:2026rxi}\footnote{
     The HM method identifies regions of parameter space \emph{not excluded} by the considered calibration data, rather than isolating a single
     best-fit point, as is the case for traditional approaches. This is achieved through a consecutive compression of the parameter-space volume
     based on a suitable implausibility measure. This yields a more realistic picture of parameter degeneracies and correlated uncertainties, which
     are represented by space-filling samples from the remaining non-implausible parameter-space volume.
     In Ref.~\cite{Iskauskas:2026rxi} the HM approach was applied to the hadronisation models accessible through \Sherpa, namely its built-in cluster
     fragmentation \textsc{Ahadic}~\cite{Chahal:2022rid}, and the Lund-string fragmentation~\cite{Andersson:1983ia} through an interface to
     \Pythia~8~\cite{Bierlich:2022pfr}.}.
Variations within the best-fit regions of the model-parameter space resulting from the tuning can be used to estimate model-inherent uncertainties or to study
the sensitivity of physical observables on the parameters. In practice, however, producing dedicated simulation runs for large multi-dimensional
parameter-space samples is computationally prohibitive; this frequently results in a rather superficial assessment of such parametric uncertainties
obtained, e.g., from direct comparison of two MCEGs or of different tunes of the same MCEG.

This paper aims to address this problem. In particular, we introduce and validate a method that offers a promising route to accelerated non-perturbative
model explorations, parameter-sensitivity studies and uncertainty estimation.  It allows the propagation of parameter-uncertainty variations, e.g.\ from a
wave of HM, without the need for multiple event-simulation runs.  While initial steps towards reweighting techniques for hadronisation models have recently
been reported by the MLHad collaboration~\cite{Bierlich:2023fmh,Assi:2025gog,Butter:2025wxn} for the Lund-string fragmentation in \Pythia, such
implementations do not yet exist for Multiple Parton Interactions (MPI) and Colour Reconnection (CR) models.
In this work, we present the first implementation of on-the-fly reweighting for MPI and CR parameters within \Sherpa, extending the scope of existing
perturbative reweighting frameworks to all key non-perturbative components of event generation. The resulting approach provides \emph{in-situ derived}
alternative event weights corresponding to specified variations of model parameters.  This enables the efficient and robust exploration of the parameter
space.  Our new method significantly accelerates the generation of alternative generator predictions, reduces the storage requirements for the samples, and
alleviates post-processing efforts, for example event analyses or detector simulations,  because, effectively, only one event sample is generated,
where vectors of relative event-by-event weights capture the effect of the parameter variations.

Our paper is structured as follows: In Section~\ref{sec:review} we review the \Sherpa MPI and CR models and outline the reweighting approach.
In Section~\ref{sec:tuning} we demonstrate the capabilities of this new functionality by performing a tuning of the \Sherpa MPI and CR models
to LHC data at $\sqrt{s}=7\,\text{TeV}$ using {\sc Apprentice} in conjunction with the reweighting framework, covering a broad parameter space with a
single dedicated run. We furthermore illustrate the application of the reweighting approach to assess the energy
extrapolation of dimensionful model parameters. We conclude in Section~\ref{sec:conclusions} with a summary and an
outlook.

\section{Setting the scene}\label{sec:review}

Following the standard collinear-factorisation approach for modelling hadron--hadron
collisions, single $2\to n$ partonic-scattering processes are considered, with one
initial-state parton extracted from each of the incident hadrons. In MCEGs this primary or
hard-process configuration is then subjected to parton-shower evolution and hadronisation.
At low scales of a few $\Lambda_{\rm QCD}$, the partons emerging from the
shower will fragment into primary hadrons, some of which will decay further.  This process has been
studied at exquisite accuracy at \LEP, mainly at the $Z$-pole, and, as a consequence, the non-perturbative
parameters describing hadronisation are usually calibrated to these data, cf.\ for
example~\cite{Skands:2014pea,Bellm:2019owc,Chahal:2022rid,Knobbe:2023njd,Iskauskas:2026rxi}. However,
more recently, data from HERA, flavour factories, and the LHC have also been included in
fragmentation-model tuning~\cite{Knobbe:2023ehi,Gieseke:2025mcy}.

When attempting to use this to describe final-state observables at hadron colliders, there
is striking evidence for additional effects.  There are certainly two contributions
that can be distinguished. Firstly, the incident partons can carry a small but non-vanishing
transverse momentum relative to the beam axis, often called \emph{intrinsic}-$k_T$. Secondly,
the remnants of the incident hadrons can interact, resulting in additional final-state hadronic activity, originally
referred to as \emph{jet-pedestal effect}~\cite{UA1:1983hhd}, nowadays known as the
Underlying Event (UE)~\cite{Caines:2009iy,CDF:2001onq,CDF:2010pdo,ATLAS:2010kmf,CMS:2011qzf,ALICE:2011ac}.
We here focus explicitly on the second component. The modelling of the UE is treated differently in
MCEGs~\cite{Buckley:2011ms,Campbell:2022qmc}.
To describe the UE, Multiple Parton Interactions (MPIs), i.e.\ secondary $2\to 2$ QCD scatters,
are introduced, that also undergo parton-shower evolution and hadronisation. In Section~\ref{sec:MPIreweighting}
we will give a brief review of the model and illustrate how its inherent parameters can be reweighted on-the-fly.

The parton-shower evolution of the primary and secondary scatterings is accomplished in the
large-$N_c$ approximation\footnote{
            To be precise, typically not the strict 't Hooft limit~\cite{tHooft:1974pnl}
            ($N_c\to \infty$ with $\alpha_sN_c=\text{const.}$) is used. Instead, formally
            subleading $g\to q\bar{q}$ splittings are included and the QCD Casimirs appearing in the
            splitting functions remain at their proper $N_c=3$ values, i.e.\ $C_A=3$ and $C_F=4/3$, see for
            instance~\cite{Schumann:2007mg}.}.
This results in planar colour flows only, where each quark and gluon has well defined colour partner(s),
facilitating the transition from partons to colour-neutral hadronic states (i.e., the clusters in \textsc{Ahadic})
at lower scales~\cite{Amati:1979fg}.  While, typically, the hard process is evaluated using $N_c=3$,
the above treatment requires the (probabilistic) assignment of a planar colour flow that seeds the subsequent
shower evolution, in which subleading colour configurations in the splitting processes are neglected.

In the presence of secondary scatterings, neither the colour structure of the beam remnants
nor the colour correlations between the MPI scatters and the hard process are uniquely defined.
To at least approximately include subleading perturbative configurations, to effectively model the
exchange of non-perturbative emissions, and to account for alternative beam-remnant colour assignments,
MCEGs include the option of Colour Reconnections (CRs), i.e.\ the probabilistic rearrangement of
colour--anti-colour partners. In Section~\ref{sec:CRreweighting} we detail the model used in \Sherpa
and how its parameters can be reweighted on-the-fly.

The subtle interplay of the models for MPIs and CRs in affecting the final-state hadronic activity
implies that they cannot be calibrated with experimental data in isolation. Accordingly, in
Section~\ref{sec:tuning} we will consider two scenarios for the tuning, one with MPI only and one with
both MPI and CR enabled.

\subsection{Reweighting the Underlying Event in \Sherpa}\label{sec:MPIreweighting}
A comprehensive understanding of the model's parameter dependences is essential to achieve the correct
reweighting of the MPI model across a varied parameter space. The following is a brief review of the MPI model in
\Sherpa, which essentially re-implements the Sj\"ostrand--van-Zijl (SZ)
model~\cite{Sjostrand:1987su}.

We begin by specifying the geometrical configuration for the incident beam particles participating in the
MPIs in \Sherpa.  The hadronic matter distribution of each incident hadron, parametrised by form factors $\rho(r)$, is
assumed to be spherically symmetric and independent of parton flavour or longitudinal momentum fraction~\footnote{
        In fact, there are options within \Pythia~\cite{Corke:2011yy} and, similar, in \Sherpa for the form factors
        to depend on both distance $r$ and longitudinal momentum fraction $x$ of the partons within the hadron.
        We will ignore this modelling idea in the following.
}.
In parallel to the SZ model, \Sherpa models this as a superposition of two Gaussian distributions,
\begin{equation}
  \rho(r) = \sum_{i=1}^{2} \frac{\alpha_i}{(\sqrt{\pi} R_i)^3} \exp\left(-\frac{r^2}{R_i^2}\right)\,,
\end{equation}
where $R_i/\sqrt{2}$ denote the Gaussian widths (w.l.o.g.\ we assume $R_1 \leq R_2$) and $\alpha_i$ are the
fractional weights satisfying $\alpha_1 + \alpha_2 = 1$. The normalisation is chosen such that
$\int {\rm d}^3x\, \rho(r) = 1$.

Given these single-hadron matter profiles, the geometry of a hadron--hadron collision is encoded in the
time-integrated spatial overlap of the two distributions. For two hadrons with density profiles $\rho_1(r)$ and $\rho_2(r)$
moving along the $\pm z$-direction separated by the transverse impact parameter $b$, the overlap function reads
\begin{align}
  O(b) \;=\; \int {\rm d}t \int {\rm d}^3x \rho_1(x, y, z) \rho_2(x, y, z - \sqrt{b^2 + t^2})
  \;=\; \frac{1}{\pi} \sum_{i,j=1}^{2} \frac{\alpha_i \alpha_j}{R_i^2 + R_j^2} \exp\left(-\frac{b^2}{R_i^2 + R_j^2}\right)\,.
  \label{eq:overlap}
\end{align}
The overlap function is normalised to $\int {\rm d}^2b\, O(b) = 1$ and thus quantifies the probability density for
parton encounters at a given impact parameter $b$.  It enters the calculation of the perturbative QCD cross section
for parton--parton interactions, as
\begin{equation}
  \sigma_{ij \to kl} = \int {\rm d}p_\perp^2 \, {\rm d}^2b \, {\rm d}\xi_1 \, {\rm d}\xi_2 \, O(b) \, f_i(\xi_1, \mu_F^2) \, f_j(\xi_2, \mu_F^2) \, \frac{{\rm d}\hat{\sigma}_{ij \to kl}(\mu^2_R)}{{\rm d}p_\perp^2}\,.
\end{equation}
The factorisation and renormalisation scales are set to $\mu_{F,R}^2 = c_{F,R}^2\, p_\perp^2$, with prefactors $c_{F,R}$ and $p_\perp$ denoting
the transverse momentum of the partonic scattering.  To avoid the divergence at $p_\perp \to 0$, a minimum transverse momentum cutoff
$p_{\perp,\min}$ is introduced, below which no perturbative scatterings are considered.  Therefore, the total hard scattering cross section
is given as a sum over all contributing partonic channels
\begin{equation}
  \sigma_{\rm hard}(p_{\perp,\min}) = \sum_{ijkl} \int_{p_{\perp,\min}^2}^{\infty} {\rm d}p_\perp^2 \int {\rm d}^2b \, {\rm d}\xi_1 \, {\rm d}\xi_2 \, O(b) \, f_i(\xi_1, \mu_F^2) \, f_j(\xi_2, \mu_F^2) \, \frac{{\rm d}\hat{\sigma}_{ij \to kl}(\mu^2_R)}{{\rm d}p_\perp^2}\,.
  \label{eq:sigma_hard}
\end{equation}
While the sharp cutoff removes the formal divergence, it enforces an abrupt and somewhat artificial separation between “hard”
and “soft” scatterings. To obtain a smoother transition for small transverse momenta, the partonic cross section is further regularised by a damping factor
\begin{equation}
  \frac{p_\perp^4}{(p_\perp^2 + p_{\perp,0}^2)^2}\,,
\end{equation}
and the strong coupling $\alpha_s(\mu_R^2)$ is evaluated at an effective scale $\mu_R^2 = c_R^2(p_\perp^2 + p_{\perp,0}^2)$.

The average number of hard interactions above the transverse-momentum threshold $p_{\perp,\min}$ is given by the ratio of the
perturbative cross section \eqref{eq:sigma_hard} and the non-diffractive hadronic cross section,
\begin{equation}
  \langle n(p_\perp > p_{\perp,\min}) \rangle = \frac{\sigma_{\rm hard}(p_{\perp,\min}, p_{\perp,0})}{\sigma_{\rm ND}}\,.
\end{equation}
Assuming independent parton–parton scatterings, the number of interactions at a given impact parameter follows a Poisson distribution
with mean $\bar{n}(b, p_{\perp,\min})$. Following this, the probability for two passing hadrons to undergo a collision (i.e., at least one
parton--parton interaction above $p_{\perp,\min}$) is obtained from Poissonian statistics as
\begin{equation}
  P_{\rm int}(b, p_{\perp,\min}) = 1 - \exp\bigl(-\bar{n}(b, p_{\perp,\min})\bigr)\,,
  \label{eq:pint}
\end{equation}
and the average number of interactions at a given $b$ in a non-diffractive event is therefore
\begin{equation}
  \langle n(b, p_{\perp,\min}) \rangle = \frac{\bar{n}(b, p_{\perp,\min})}{P_{\rm int}(b, p_{\perp,\min})}\,.
\end{equation}
Integrating over all impact parameters then recovers the overall average number of interactions
\begin{equation}
  \langle n \rangle = \frac{\int {\rm d}^2b \, \langle n(b, p_{\perp,\min}) \rangle P_{\rm int}(b, p_{\perp,\min})}{\int {\rm d}^2b \, P_{\rm int}(b, p_{\perp,\min})} = \frac{\int {\rm d}^2b \, \bar{n}(b, p_{\perp,\min})}{\int {\rm d}^2b \, P_{\rm int}(b, p_{\perp,\min})} = \frac{\sigma_{\rm hard}}{\sigma_{\rm ND}}\,.
\end{equation}
Comparison with the hard cross section (\ref{eq:sigma_hard}) suggests that the impact-parameter-dependent MPI rate is given by
\begin{equation}
  \bar{n}(b, p_{\perp,\min}) = O(b) \,\sum_{ijkl} \int_{p_{\perp,\min}^2}^{\infty} {\rm d}p_\perp^2 \, {\rm d}\xi_1 \, {\rm d}\xi_2 \, f_i(\xi_1, \mu_F^2) \, f_j(\xi_2, \mu_F^2) \, \frac{{\rm d}\hat{\sigma}_{ij \to kl}(\mu^2_R)}{{\rm d}p_\perp^2}\,.
\end{equation}
The integrated interaction probability must therefore equal the non-diffractive cross section
\begin{equation}
  \int {\rm d}^2b \, P_{\rm int}(b) = \int {\rm d}b\, 2\pi b \left[1 - \exp\bigl(-\bar{n}(b, p_{\perp,\min})\bigr)\right] = \sigma_{\rm ND}\,.
  \label{eq:sigma_nd}
\end{equation}
This expression establishes the connection between the matter-distribution parameters and the observable MPI rates.
In order to satisfy this relation, the matter-distribution widths are tuned via a scaling factor $R^{\rm eff}_i = k \cdot R_i$~\footnote{
   In contrast to the original SZ model, this interpretation affords some physical meaning to the radii of the
   matter distribution.  It is satisfying to note that, after rescaling, the resulting $R^{\rm eff}_i$ are of
   the order of 0.1-1 fm.
}.

For the actual event simulation, multiple parton scatters are generated in an ordered sequence of decreasing $p_\perp$.  Similar to the treatment
of multiple emissions in parton showers, a Sudakov-like form factor encodes the suppression of additional interactions.
The exponential probability for no scattering between two successive transverse-momentum scales is thus given by
\begin{equation}
  \Delta(b) = \exp\left(-\int_{p_{\perp,\text{next}}^2}^{p_{\perp,\text{prev}}^2} 
  \mathrm{d}p_\perp^2 \int \mathrm{d}\xi_1 \,\mathrm{d}\xi_2 \, O(b) 
  \sum_{ijkl} f_i(\xi_1, \mu_F^2)\, f_j(\xi_2, \mu_F^2)\, 
  \frac{\mathrm{d}\hat{\sigma}_{ij\to kl}(\mu_R^2)}{\mathrm{d}p_\perp^2}\right)\,.
\end{equation}
The generation of MPIs proceeds via a downward evolution in $p_\perp$, starting from a maximum scale set by the hard process, by solving the
Sudakov form factor equation for the next lower $p_\perp$ scale. In practice, an integrable overestimator is used to propose a new trial
$p_\perp$, which is then accepted or rejected based on the true integrand, called the Sudakov Veto Algorithm (SVA)~\cite{Seymour:1994df}.
This procedure is iterated until the transverse momentum falls below the minimum cutoff $p_{\perp,\min}$, at which point no further hard interactions
are generated~\footnote{
    Each (non-perturbative) event starts with a signal process, calculated and described through traditional collinear factorisation and
    perturbative expansion, which also induces a hard scale $p_\perp^{\rm hard}$.  In a first step, the algorithm generates trial pairs of
    impact parameter $b$, sampled according to ${\rm d}^2b\,O(b)$, and the scale of the hardest MPI scatter $p_\perp^{\rm (MPI)}$, obtained
    from $\Delta(b)$ with the upper scale the c.m.-energy of the colliding hadrons.  If $p_\perp^{\rm (MPI)}>p_\perp^{\rm hard}$ a new trial pair
    is generated; conversely, if $p_\perp^{\rm (MPI)}<p_\perp^{\rm hard}$, the configuration is accepted, the impact parameter is fixed, and the
    generation of MPIs proceeds with the first additional scatter fixed.  In contrast, {\em inclusive} (i.e.\ Minimum Bias) events start
    by fixing the impact parameter according to $P_{\rm int}(b)$ and selecting the first scatter freely from $\Delta(b)$ with a starting
    scale equivalent to the hadronic c.m.\ energy. The difference between $O(b)$ and $P_{\rm int}(b)$ is shown in Fig.~\ref{fig:overlap_and_pint};
    clearly, the overlap $O(b)$ is described by a significantly sharper distribution than $P_{\rm int}(b)$, resulting in the underlying
    event usually driven by smaller impact parameters $b$ and, correspondingly, a larger partonic activity compared to minimum-bias
    events.  In this publication, however, we have not used minimum-bias data and will instead aim for a combined fit of the model with
    underlying-event and minimum-bias data in future work.
}.

This implementation suggests to split the reweighting factor for MPI model-parameter variations into two components: the first factor
$w_b$ accounts for a modified impact-parameter sampling and the second factor $w_{\rm Sudakov}$ captures the variations of the Sudakov
factors at a given impact parameter (and thus a varied MPI multiplicity). Following this, each event will receive a total weight of
\begin{equation}
w_i = (w_b)_i \cdot (w_{\rm Sudakov})_i
\end{equation}
for the parameter variation $i$. The computation of the impact-parameter reweighting factor $w_b$ is straightforward. Given an
original impact parameter $b$ sampled from the overlap function $O(b; \vec{p}_0)$ with parameter set $\vec{p}_0$, the modified weight
for a new parameter set $\vec{p}_i$ is given by the ratio of the overlap functions evaluated at the same impact parameter
\begin{equation}
  (w_b)_i = \frac{O(b; \vec{p}_i)}{O(b; \vec{p}_0)}\,.
  \label{eq:wb}
\end{equation}

For the Sudakov reweighting factor $w_{\rm Sudakov}$, the procedure is more involved. As the overestimator used in the SVA depends on the
MPI parameters, using the maximum overestimator across all variations is necessary to ensure the validity of the SVA for all parameter
sets.  With this rescaling, the reweighting factor can be constructed from each accepted and rejected secondary scattering in the
event-generation sequence. For each proposed trial $p_\perp$, the acceptance probability $\mathcal{P}^{\rm acc}$ of a parameter variation
is given by the ratio of the true integrand for the varied and original parameter sets multiplied by the acceptance probability of the
original parameter set
\begin{align}
  \mathcal{P}^{\rm acc}_i(p_\perp) =&\: \frac{O(b; \vec{p}_i) \sum_{ijkl} f_i(\xi_1, \mu_F^2) \, f_j(\xi_2, \mu_F^2) \, \frac{{\rm d}\hat{\sigma}_{ij \to kl}}{{\rm d}p_\perp^2}(\mu^2_R; \vec{p}_i)}{{\rm overestimator}(p_\perp)} \nonumber\\
   =&\: \mathcal{P}^{\rm acc}_0(p_\perp) \cdot \frac{O(b; \vec{p}_i) \sum_{ijkl} f_i(\xi_1, \mu_F^2) \, f_j(\xi_2, \mu_F^2) \, \frac{{\rm d}\hat{\sigma}_{ij \to kl}}{{\rm d}p_\perp^2}(\mu^2_R; \vec{p}_i)}{O(b; \vec{p}_0) \sum_{ijkl} f_i(\xi_1, \mu_F^2) \, f_j(\xi_2, \mu_F^2) \, \frac{{\rm d}\hat{\sigma}_{ij \to kl}}{{\rm d}p_\perp^2}(\mu^2_R; \vec{p}_0)}\,.
\end{align}
Consequently, for an accepted partonic scatter at scale $p_\perp$ the Sudakov reweighting factor receives a contribution of
\begin{equation}
  q_i^{\rm acc}(p_\perp) = \frac{\mathcal{P}^{\rm acc}_i(p_\perp)}{\mathcal{P}^{\rm acc}_0(p_\perp)}\,,
\end{equation}
and for each rejected interaction the contribution is given by
\begin{equation}
  q_i^{\rm rej}(p_\perp) = \frac{1 - \mathcal{P}^{\rm acc}_i(p_\perp)}{1 - \mathcal{P}^{\rm acc}_0(p_\perp)}\,.
\end{equation}
For an event with several accepted and rejected secondary scatterings, the total Sudakov reweighting factor is given by the product of
all contributions.

Reweighting is supported for all key parameters of the MPI model introduced above. Specifically, these include the hadronic-matter
distribution parameters $R_1$, $R_2$, and $\alpha_1$ that directly influence the overlap function $O(b)$. The non-diffractive cross section
can be rescaled by a normalisation factor $\sigma_{\rm ND}^{\rm norm}$ to increase or decrease the MPI rate, which through the
constraint (\ref{eq:sigma_nd}) induces a corresponding rescaling $k$ of the effective Gaussian widths and thereby modifies the overlap.

Variations in these four parameters consequently affect both the impact-parameter distribution and the Sudakov factors through
variations of $O(b)$. Additionally, the model supports variations of the transverse-momentum scales $p_{\perp,0}$ (controlling the low-$p_\perp$
spectrum shape) and $p_{\perp,\min}$ (setting the phase-space boundary of $\sigma_{\rm hard}$), with variations in either of these two parameters
affecting $O(b)$ through a change in the $k$ factor, and directly modifying the Sudakov suppression factors through the so
altered MPI acceptance probabilities.

A technical caveat applies to $p_{\perp,\min}$.  For this parameter only upward variations (i.e., increasing the $p_\perp$ threshold to
exclude softer MPIs) are possible, since events generated with an original higher cutoff simply do not fill the phase space and hence cannot be
reweighted.  However, for a trial $p_\perp$ below the varied cutoff, the acceptance probability is 0, i.e.
\begin{equation}
  \mathcal{P}^{\rm acc}_i(p_\perp < (p_{\perp,\min})_i) = 0\,,
  \label{eq:ptmin_acc}
\end{equation}
ensuring that events with an accepted MPI below the varied cutoff receive a total weight of 0.

Inspired by the Donnachie--Landshoff model for the scaling of the total hadronic cross sections~\cite{Donnachie:1992ny},
the energy dependence of $p_{\perp,0}$ and $p_{\perp,\min}$ is parametrised as
\begin{equation}\label{eq:pts}
  p_{\perp,0}(s) = p_{\perp,0}^{\rm ref} \left(\frac{s}{s_{\rm ref}}\right)^{\eta}\,, \quad p_{\perp,\min}(s) = p_{\perp,\min}^{\rm ref} \left(\frac{s}{s_{\rm ref}}\right)^{\eta}\,,
\end{equation}
where $s_{\rm ref}$ is a reference collision-energy scale, and $\eta$ controls the energy scaling.
Variations in $\eta$ are also supported by the reweighting procedure, as this simply amounts to variations in $p_{\perp,0}$ and $p_{\perp,\min}$.

Once the generation of MPIs and subsequent showers have terminated for a single event, the resulting parton
configuration must be hadronised, introducing an observable dependence on how colour quantum numbers (in the $N_c\to\infty$ limit)
are distributed over the partons.  Realising that for each scatter such colour assignments are more or less straightforward,
the relevant question therefore concerns the distribution of colours between the individual parton scatters and, ultimately,
with the spectator partons inside the beam remnants.

In \Sherpa, this problem is solved by assigning colours iteratively.  Initially, each individual scattering event undergoes
parton showering with the corresponding colour assignments.  Once the shower terminates the colours of the two resulting
initial-state partons will be redefined such that they are colour-connected to the two initial-state partons of the previous
scatter in each hadron.  As a consequence this leads to colour-connections between the final-state partons of subsequent
MPIs.  As a final step, \Sherpa assigns spectator partons -- and, in particular, diquarks in the case of incident baryons --
in each of the beam remnants to ensure their correct flavour quantum numbers.  These spectator partons will also carry the
``trailing'' disconnected colours.

\begin{table}
\centering
  \caption{MPI model parameters in \Sherpa that can be reweighted.\label{tab:MPIparams}}
  \begin{tabular}{lll}
    \hline
    Parameter & Description & Name (in run card) \\
    \hline
    $R_1$ & Gaussian width of core matter distribution & \texttt{MATTER\_RADIUS\_1} \\
    $R_2$ & Gaussian width of halo matter distribution & \texttt{MATTER\_RADIUS\_2} \\
    $\alpha_1$ & Fractional weight of first Gaussian & \texttt{MATTER\_FRACTION\_1} \\
    $\sigma_{\rm ND}^{\rm norm}$ & Multiplicative factor for $\sigma_{\rm ND}$ & \texttt{SIGMA\_ND\_NORM} \\
    $p_{\perp,0}^{\rm ref}\,[\text{GeV}]$ & IR regularisation scale & \texttt{PT\_0(ref)} \\
    $p_{\perp,\min}^{\rm ref}\,[\text{GeV}]$ & IR cutoff scale  & \texttt{PT\_Min(ref)} \\
    $\eta$ & Energy scaling exponent & \texttt{Eta}\\
    \hline
  \end{tabular}
\end{table}

\subsection{Reweighting Colour Reconnections in \Sherpa}\label{sec:CRreweighting}
The \Sherpa CR model is invoked after the parton-shower evolution of the hard process
and the secondary scatterings, but prior to the non-perturbative splitting of all final-state gluons and the
formation of colour-neutral clusters in its cluster-hadronisation model~\cite{Chahal:2022rid}. It assumes a
colour flow in the large-$N_c$ limit with a total of $N$ colour--anti-colour pairs and performs a sequence
of reconnection attempts, each based on a random choice of two colour lines. For every candidate reconnection,
the model evaluates a distance measure in momentum space.

\Sherpa supports two such distance measures between two partons $i$ and $j$ in momentum space, given by
\begin{equation}
 d_{ij} =
\begin{cases}
\log\left[1 + \dfrac{(p_i \cdot p_j - m_i m_j)}{Q_0^2}\right] & \text{(logarithmic)} \\
\left(\dfrac{p_i \cdot p_j - m_i m_j}{\langle d\rangle}\right)^{\kappa} & \text{(power)}
\end{cases}
\end{equation}
where $p_i$ and $p_j$ are the parton four-momenta, and $m_i$ and $m_j$ their non-perturbative constituent masses.
$Q_0$ is a tunable scale parameter, $\kappa$ a tunable exponent, and $\langle d\rangle$ the average distance over all pairs.
If either $i$ or $j$ is a gluon, only half of its momentum enters, reflecting the anticipated sharing of momentum
between its colour and anti-colour lines. By default, \Sherpa employs the logarithmic distance measure, which is also
the variant adopted for all results discussed in this paper.

In each of $N$ iterations, $N$ candidate reconnections are proposed between randomly chosen colour lines. For any trial
reconnection, $[i_1, j_1], [i_2, j_2] \rightarrow [i_1, j_2], [i_2, j_1]$, the change in total distance in momentum space is
\begin{equation}
\Delta = (d_{i_1 j_1} + d_{i_2 j_2}) - (d_{i_1 j_2} + d_{i_2 j_1})\,.
\end{equation}
Among all proposals in a given iteration, the one with the largest reduction in total distance (i.e. the largest positive $\Delta$)
is selected as the candidate reconnection. The probability to accept this reconnection is
\begin{equation}
\mathcal{P}^{\text{acc}}(\Delta) = R \left[1 - \exp(-\eta_Q^2 \, \Delta)\right]\,,
\label{eq:CRacceptance}
\end{equation}
where $\eta_Q$ and $R$ are tunable parameters.

If the reconnection is accepted, the colour lines are swapped immediately,
otherwise, the original configuration is retained.

\begin{table}
\centering
  \caption{CR model parameters in \Sherpa that can be reweighted.\label{tab:CRparams}}
  \begin{tabular}{lll}
    \hline
    Parameter & Description & Name (in run card) \\
    \hline
    $\eta_Q$ & Multiplicative factor of distance change $\Delta$ & \texttt{ETA\_Q} \\
    $R$ & Multiplicative factor of acceptance probability & \texttt{RESHUFFLE} \\
    \hline
  \end{tabular}
\end{table}

The model supports parameter variations in $\eta_Q$ and $R$. For each variation, the corresponding event weight is updated
multiplicatively at every reconnection attempt, depending on whether the candidate swap is accepted or rejected.
For accepted reconnections with distance reduction $\Delta$, the weight contribution for variation $i$ is
\begin{equation}
q_i^{\text{acc}}(\Delta) =
\frac{\mathcal{P}^{\text{acc}}(\Delta; \vec{p}_i)}{\mathcal{P}^{\text{acc}}(\Delta; \vec{p}_0)}\,,
\end{equation}
while for a rejected reconnection proposal it is
\begin{equation}
q_i^{\text{rej}}(\Delta) =
\frac{1 - \mathcal{P}^{\text{acc}}(\Delta; \vec{p}_i)}{1 - \mathcal{P}^{\text{acc}}(\Delta; \vec{p}_0)}\,.
\end{equation}
The total CR reweighting factor for a given event is obtained by multiplying all contributions from accepted and rejected reconnection attempts.
The corresponding alternative event weights, both for MPI and CR parameter variations, are accessible via the \textsc{HepMC}~\cite{Buckley:2019xhk}
event record as well as \Sherpa's internal interface to \Rivet~\cite{Bierlich:2024vqo}.
See Appendix~\ref{app} for details on the syntax and the weight-naming conventions.

\subsection{Initial Validation}

Before applying the new reweighting technique to the actual tuning of the MPI and CR parameters,
in the next Section, Sec.~\ref{sec:tuning},  we show some initial validation results.  To this end, we
consider underlying-event
simulations accompanying Drell--Yan lepton-pair production in proton--proton collisions at
$\sqrt{s}=7\,\text{TeV}$. Further details on the calculational setup will be provided in
Section~\ref{sec:setup}.

\begin{figure}[h]
    \centering
    \includegraphics[width=1.0\textwidth]{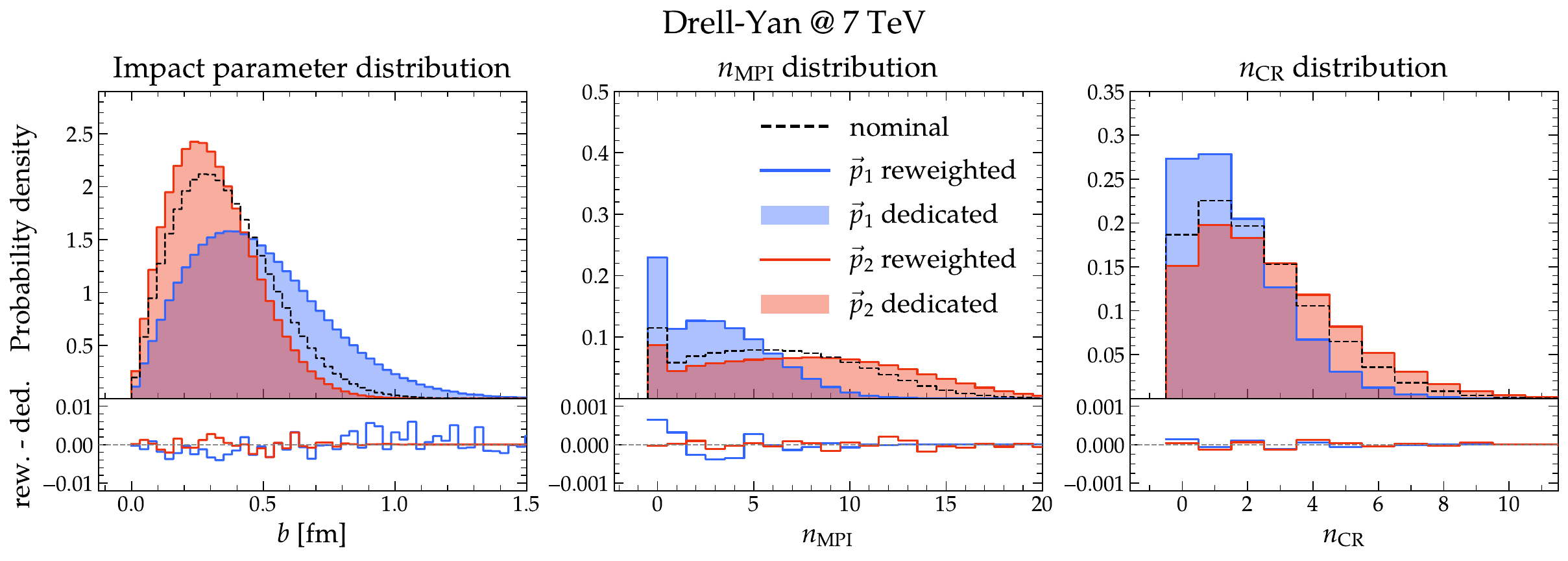}
    \caption{Distributions of the impact parameter $b$,
    the number of multiple parton interactions $n_{\rm MPI}$, and the number of colour reconnections $n_{\rm CR}$ in inclusive Drell--Yan production
    at $\sqrt{s}=7\,\text{TeV}$. The three panels display the results for variations in the MPI and CR model parameters. For the MPI variations
    (first and second panels), the parameter sets are
    $\vec{p}^{\rm\:MPI}_1 = (R_1 = 0.75~\mathrm{fm},\, \alpha_1 = 0.7,\, \sigma_{\rm ND}^{\rm norm} = 0.7,\, p_{\perp,\min}^{\rm ref} = 1.3~\mathrm{GeV})$ and
    $\vec{p}^{\rm\:MPI}_2 = (0.95~\mathrm{fm},\,0.3,\,0.3,\,1.0~\mathrm{GeV})$,
    while for the CR variation (third panel) the parameter sets are $\vec{p}^{\rm\:CR}_1 = (\eta_Q = 0.35)$ and $\vec{p}^{\rm\:CR}_2 = (0.65)$.
    The nominal parameter set is given by
    $\vec{p}_0 = (R_1 = 0.85~\mathrm{fm},\, \alpha_1 = 0.5,\, \sigma_{\rm ND}^{\rm norm} = 0.5,\, p_{\perp,\min}^{\rm ref} = 1.0~\mathrm{GeV},\, \eta_Q = 0.5)$.
    Each panel shows both the dedicated and the reweighted distributions, the latter are obtained by
    reweighting from the nominal sample, indicated by the black dashed line.
    The lower panels display the differences between both results.
    }
    \label{fig:DY_distributions}
\end{figure}

Figure~\ref{fig:DY_distributions} shows closure tests for three intrinsic model observables
that systematically validate the individual reweighting components. The first panel probes
the probability distribution of the impact-parameter $b$, testing the geometrical reweighting
factor $(w_b)_i$, cf.~Eq.~(\ref{eq:wb}). The second panel examines the MPI multiplicity $n_{\rm MPI}$,
which depends on the full MPI weight $w_i = (w_b)_i \cdot (w_{\rm Sudakov})_i$ and thus critically
tests the reweighting of the Sudakov factor. The third panel validates the CR reweighting
through the number of colour reconnections $n_{\rm CR}$. The parameter variations are specifically
designed to demonstrate successful reweighting towards both higher and lower values of each
observable (see caption for details). Dedicated samples generated directly for the varied parameter
points show excellent agreement with predictions obtained from reweighting the nominal event
sample, indicated by the dashed line.

\begin{figure}
    \centering
    \includegraphics[width=1.0\textwidth]{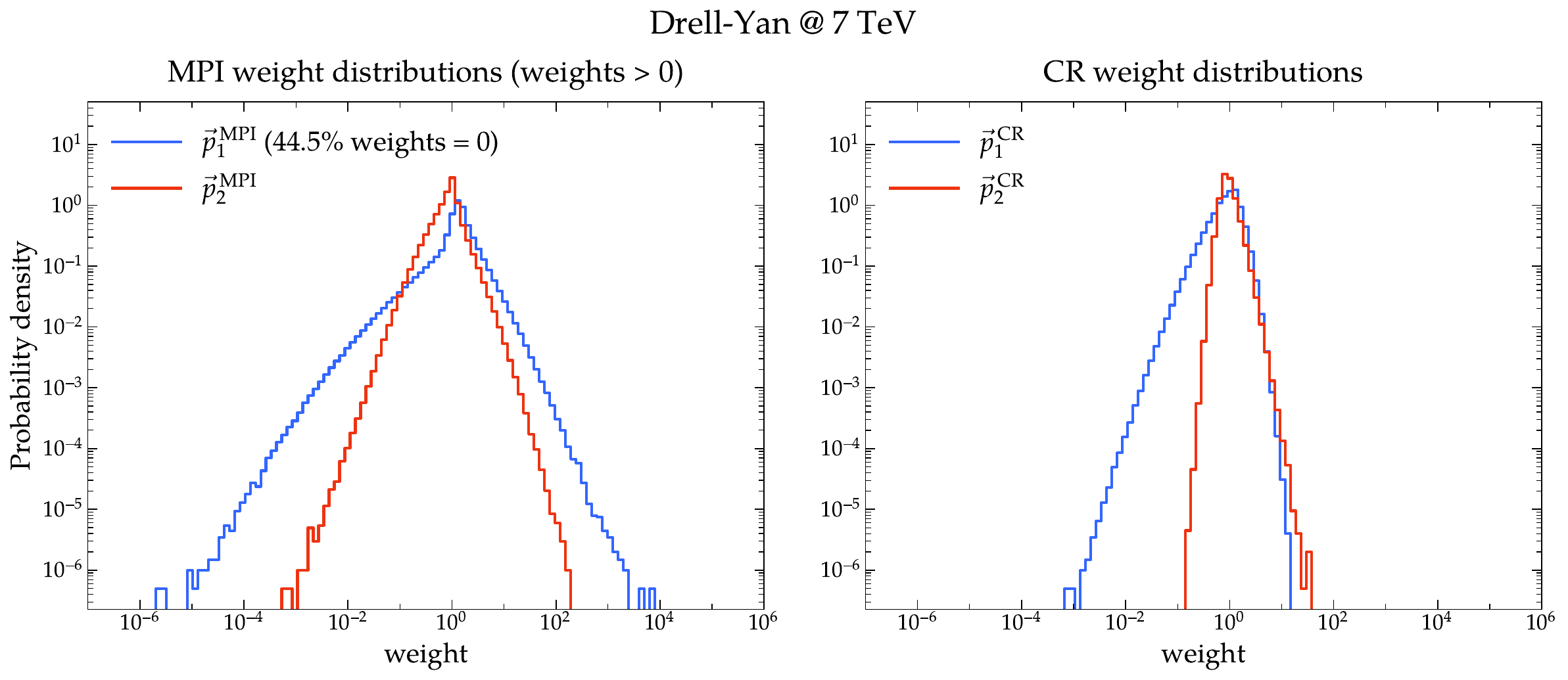}
    \caption{Distributions of the reweighting event weights obtained for the Drell--Yan process validation runs at $\sqrt{s}=7\,\text{TeV}$.
    The two panels show the weight distributions corresponding to the parameter variations used in Figure~\ref{fig:DY_distributions}.
    The left panel displays the weights for the MPI variations, and the right panel for the CR variations.}
    \label{fig:DY_weight_distribution}
\end{figure}

The corresponding distributions of reweighting weights are shown in Figure~\ref{fig:DY_weight_distribution}.
They remain well-behaved with distributions centred around unity, indicating numerical stability of the reweighting approach.
Notably, the first MPI variation ($\vec{p}^{\rm\:MPI}_1$) contains an upward variation of $p_{\perp,\min}^{\rm ref}$ from $1.0$ to $1.3~{\rm GeV}$,
which produces $\sim45\%$ zero weights due to the phase-space restriction, cf.~Eq.~(\ref{eq:ptmin_acc}). Large variations of
$p_{\perp,\min}^{\rm ref}$ thus significantly reduce the effective number of non-zero weights, a limitation that must be considered in practical
applications of the reweighting method.

While reweighting intrinsic model variables constitutes a necessary validation of the implementation,
closure tests for experimentally relevant physical observables remain essential and are presented in
the following section.

\section{Tuning methodology and results}\label{sec:tuning}

In this section we describe the tools and methods used to tune the \Sherpa MPI and CR model
parameters. As both models have an impact on the final-state hadronic activity, we consider
separate tunes for the case of just MPI without the option of CR, and MPI with CR enabled.
To facilitate a resource efficient tuning procedure, allowing us to cover a wide range of the
parameter space, we build fast surrogate models for the exact generator response to parameter
variations using the {\sc{Apprentice}}~\cite{Krishnamoorthy:2021nwv} framework. To train the
surrogate model we employ our reweighting technique introduced in Section~\ref{sec:review}.
Following a series of detailed closure tests with dedicated generator runs, we present the
physics predictions of our best-tune parameter sets. Finally, we investigate the energy scaling
of the dimensionful MPI parameters by tuning against proton--anti-proton collision data at
$\sqrt{s}=1.96$ TeV from the Tevatron and $13$ TeV LHC data.

\subsection{Generator setups and reference data selection}\label{sec:setup}

To constrain the parameters of the \Sherpa MPI and CR models we
employ proton--proton collision data taken by LHC experiments at $7$ TeV centre-of-mass energy, which
accordingly serves as reference scale for energy-dependent model parameters, cf.\ Eq.~\eqref{eq:pts}.
In particular we consider measurements of hadronic observables in Drell--Yan lepton-pair production
and pure jet final states accessible through \Rivet~\cite{Bierlich:2024vqo}. 

Our particle-level event simulations are carried out with a pre-release version of
\Sherpa-3.1~\cite{Sherpa:2024mfk} that features our revised CR modelling and the reweighting functionalities
for the MPI and CR parameters. We simulate the Drell--Yan and jet-production hard processes at
next-to-leading order accuracy in the strong coupling, using tree-level matrix elements from
\Amegic~\cite{Krauss:2001iv} and \Comix~\cite{Gleisberg:2008fv}, in conjunction with one-loop
amplitudes from \OpenLoops2~\cite{Buccioni:2019sur} using the \Collier library~\cite{Denner:2016kdg}.
The matrix elements are matched to the \Sherpa
dipole shower~\cite{Schumann:2007mg} using the \MCatNLO formalism~\cite{Hoeche:2011fd}. The input
value for the strong coupling is set consistently with the employed PDF4LHC21 proton parton-density
(PDF) set~\cite{PDF4LHCWorkingGroup:2022cjn}, accessed via \textsc{LhaPDF}~\cite{Buckley:2014ana},
to $\alpha_s(m^2_Z)=0.118$. The PDF and $\alpha_s$ settings are also used for the secondary scatterings,
for which we consider all $2\to 2$ QCD processes at tree-level accuracy, using $c_{F,R}=0.5$. The
parton-to-hadron transition is modelled with \Sherpa's cluster fragmentation~\cite{Chahal:2022rid}
using the \Sherpa-3.0 default parameters determined from tuning to \LEP data~\cite{Knobbe:2023njd}.
The decays of unstable hadrons are treated by \Sherpa's built-in hadron-decay module.

To have sensitivity to the MPI and CR modelling in the simulations we consider measurements that
focus in particular on the underlying-event contribution. For the Drell--Yan process we employ an
analysis by the ATLAS experiment studying a variety of charged-particle distributions corresponding
to events containing a $Z^0$-boson decaying to a pair of electrons or muons~\cite{ATLAS:2014yqy}
(\Rivet identifier: \texttt{ATLAS\_2014\_I1315949}).
In addition, we use a similar analysis by CMS~\cite{CMS:2012oqb} considering muonic final states
(\texttt{CMS\_2012\_I1107658}). Both analyses study charged-particle multiplicities and
transverse momenta in regions of azimuthal angle defined with respect to the $Z^0$-boson direction,
differential in the di-lepton transverse momentum. For pure jet production we consider two
dedicated underlying-event analyses, one from ATLAS~\cite{ATLAS:2014iez} (\texttt{ATLAS\_2014\_I1298811})
and one from CMS~\cite{CMS:2011qzf} (\texttt{CMS\_2011\_I916908}). Both provide measurements of
charged-particle multiplicities and momentum spectra in azimuthal regions transverse to the
leading (track-)jet in the event. Furthermore, we consider a jet-fragmentation-function measurement
from ATLAS~\cite{ATLAS:2011myc} (\texttt{ATLAS\_2011\_I929691}), a measurement of transverse
event-shape variables from CMS~\cite{CMS:2011usu} (\texttt{CMS\_2011\_I886332}), and an energy-flow
measurement at large rapidities in dijet events by CMS~\cite{CMS:2011xjg} (\texttt{CMS\_2011\_I930319}).

\subsection{Tuning methodology}\label{sec:tuning_method}

To explore the physics-model parameter space we employ the {\sc{Apprentice}}~\cite{Krishnamoorthy:2021nwv}
tuning tool. For all bins of the considered observable histograms {\sc{Apprentice}} constructs a
multivariate rational surrogate model $R(\vec{p})=P(\vec{p})/Q(\vec{p})$ that aims to capture the
simulator response on variations of the model parameters $\vec{p}$. For the tuning problem at hand we
found a satisfying approximation quality for a simple polynomial function, i.e.\ $P(\vec{p})$ of degree five
and $Q(\vec{p})$ of degree zero. The statistical errors of the MC predictions are accounted for by constructing
additional surrogate models of the same form for the uncertainties in each histogram bin.

We consider two scenarios for the tuning:

\begin{enumerate}
\item[(i)] \textbf{MPI-CRoff}: The option to probabilistically rearrange the colour assignments of partons is
disabled, the tuning comprises the four MPI model parameters listed in Table~\ref{tab:paramstunerange}, where
we also quote the initial parameter ranges.
\item[(ii)] \textbf{MPI-CRon}: CR is enabled and the tuning furthermore includes the $\eta_Q$ parameter,
cf.~Table~\ref{tab:paramstunerange} for the considered parameter ranges.
\end{enumerate}

\begin{table}
\centering
  \caption{MPI and CR model parameters and their values and maximal ranges considered in the
  \textbf{MPI-CRoff} and \textbf{MPI-CRon} tunes, respectively.\label{tab:paramstunerange}}
  \begin{tabular}{lcc}
    \hline
    & \multicolumn{2}{c}{value/tune range}\\
    Parameter & \textbf{MPI-CRoff} & \textbf{MPI-CRon}\\
    \hline
    $R_1\,[\text{fm}]$ & \multicolumn{2}{c}{$[0.75,0.95]$}\\
    $R_2\,[\text{fm}]$ & \multicolumn{2}{c}{$1.0$}\\
    $\alpha_1$ & \multicolumn{2}{c}{$[0.30,0.70]$} \\
    $\sigma_{\rm ND}^{\rm norm}$ & \multicolumn{2}{c}{$[0.30,0.70]$} \\
    $p_{\perp,0}^{\rm ref}\,[\text{GeV}]$ & \multicolumn{2}{c}{$2.05$} \\
    $p_{\perp,\min}^{\rm ref}\,[\text{GeV}]$ & \multicolumn{2}{c}{$[1.00,1.30]$} \\
    $c_R$ & \multicolumn{2}{c}{$0.5$} \\
    $c_F$ & \multicolumn{2}{c}{$0.5$} \\
    $\eta_Q$ & -- & $[0.35,0.65]$ \\
    $R$ & -- & $1/9$ \\
    $Q_0\,[\text{GeV}]$ & -- & $1.0$ \\
    \hline
  \end{tabular}
\end{table}

To cover the available model space, we generate $500$ random parameter configurations uniformly
distributed over the four- and five-dimensional parameter hyperrectangle, respectively. Instead of performing
dedicated \Sherpa simulations for each of these points, we employ the reweighting method described in
Sections~\ref{sec:MPIreweighting} and \ref{sec:CRreweighting}. As the nominal parameter set, we use 
$$\vec{p}_0 = \left(R_1=0.85\,\text{fm}, \alpha_1=0.5, \sigma_{\rm ND}^{\rm norm}=0.5, p_{\perp,\min}^{\rm ref}=1.0\,\text{GeV}\right)\,,$$
which is extended by the CR parameter $\eta_Q=0.5$ \footnote{
    Reweighting is supported for both $\eta_Q$ and $R$, but expanding the acceptance probability~(\ref{eq:CRacceptance}) to first order 
    in $\eta_Q^2\Delta$ gives $\mathcal{P}^{\text{acc}}(\Delta)\approx R\,\eta_Q^2\,\Delta$, so that the two are largely 
    degenerate. We therefore fix $R=1/9$ and vary $\eta_Q$ alone.
} for the \textbf{MPI-CRon} scenario.
Accordingly, from these nominal samples we produce predictions for the $500$ parameter points,
respectively, through reweighting. Compared to the time needed to generate $500$ dedicated samples,
this approach reduces the compute time by about a factor $250$. We briefly discuss the scaling
behaviour of the reweighting method in Appendix~\ref{app2}, see in particular Fig.~\ref{fig:runtime}. 

As several analyses targeting different processes enter the tune simultaneously, they can contribute very 
differently to the objective function. To avoid any single process dominating the goodness-of-fit (GOF), and thereby 
biasing the resulting parameters, we rescale the per-observable weights of each process group (Drell--Yan/jets) by a common factor 
such that all groups contribute comparably. These factors are fixed from the minimal GOF of separate fits to the individual processes.

Before performing the parameter optimisation, we validate the reweighting procedure by means of dedicated 
closure tests. For each model parameter individually, we compare predictions obtained by reweighting the 
nominal sample $\vec{p}_0$ against dedicated \Sherpa runs generated directly for the varied parameter point, while 
keeping all other parameters fixed at their nominal values. The parameter variations are chosen to cover the respective tuning 
ranges listed in Table~\ref{tab:paramstunerange}, probing the boundaries of the parameter hyperrectangle along each individual
axis\footnote{
    The variations of parameters $R_1$ and $\alpha_1$ are combined into variations of the matter distribution $\rho$. 
    For the closure tests, we consider a compact core profile ($R_1=0.75\,\text{fm}$, $\alpha_1=0.7$) and a more diffuse profile ($R_1=0.95\,\text{fm}$, $\alpha_1=0.3$).}.

\begin{figure}
    \centering
    \includegraphics[width=1.\textwidth]{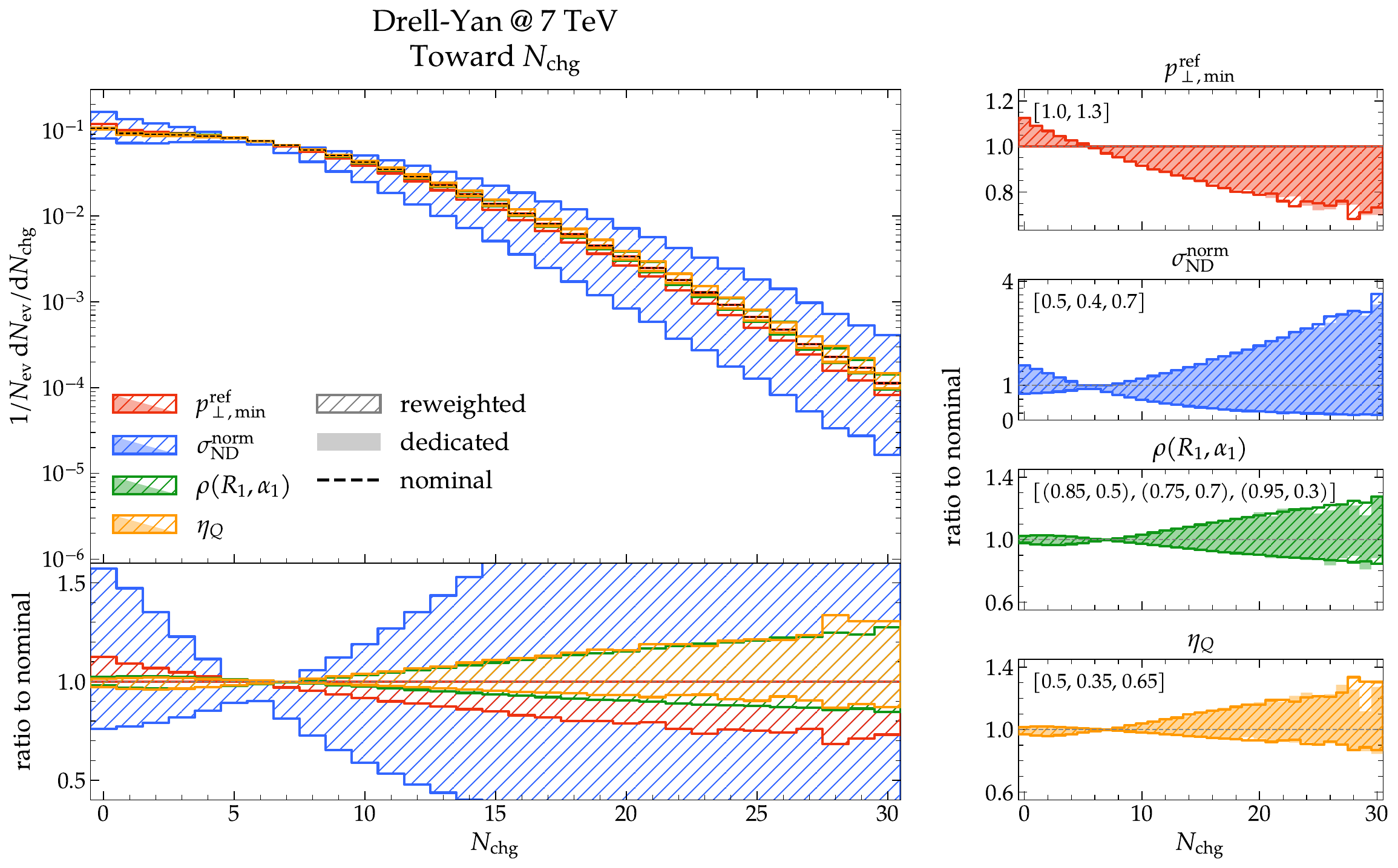}
    \caption{The number of charged particles $N_{\rm ch}$ in the toward region in Drell--Yan events at $\sqrt{s}=7\,\text{TeV}$,
    with independent variations of the MPI parameters $p_{\perp,\min}^{\rm ref}$ (red), $\sigma_{\rm ND}^{\rm norm}$ (blue) 
    and $\rho(R_1, \alpha_1)$ (green) and the CR parameter $\eta_Q$ (orange). The left panel shows the variations obtained from the reweighting procedure, with the nominal 
    parameter set $\vec{p}_0$ described in Section~\ref{sec:tuning}. The panels on the right compare the reweighting-based variations 
    to those obtained from dedicated generator runs at the varied parameter points. Parameter values are given in the notation [nominal, variation 1, (variation 2)].}
    \label{fig:closure_DY_CR_on}
\end{figure}

\begin{figure}
    \centering
    \includegraphics[width=1.\textwidth]{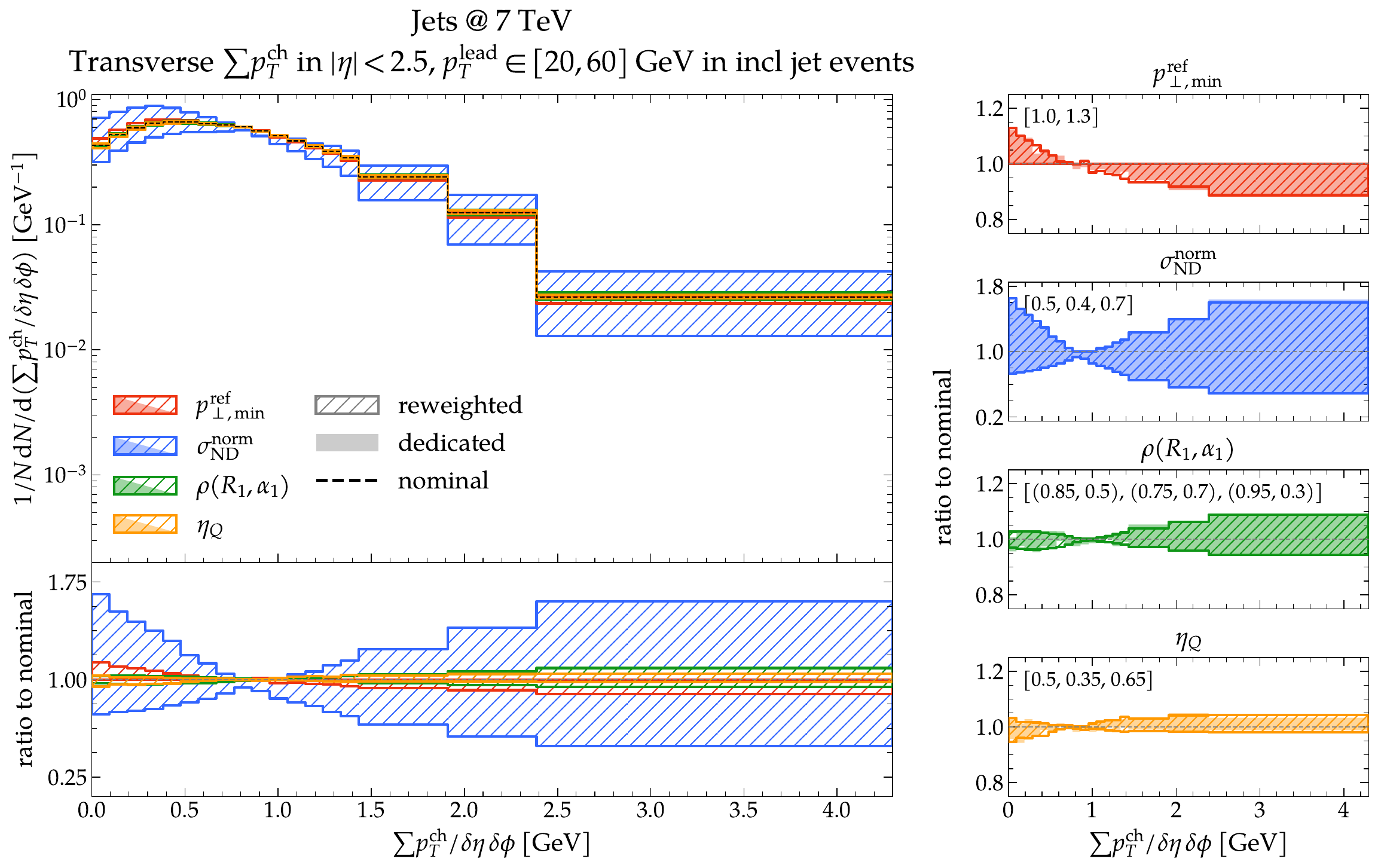}
    \caption{Sum of transverse momenta of charged particles in the transverse region in inclusive jet events at $\sqrt{s}=7\,\text{TeV}$,
    with independent variations of the MPI parameters $p_{\perp,\min}^{\rm ref}$ (red), $\sigma_{\rm ND}^{\rm norm}$ (blue) 
    and $\rho(R_1, \alpha_1)$ (green) and the CR parameter $\eta_Q$ (orange). The left panel shows the variations obtained from the reweighting procedure, with the nominal 
    parameter set $\vec{p}_0$ described in Section~\ref{sec:tuning}. The panels on the right compare the reweighting-based variations 
    to those obtained from dedicated generator runs at the varied parameter points. Parameter values are given in the notation [nominal, variation 1, (variation 2)].}
    \label{fig:closure_Jets_CR_on}
\end{figure}

Figures~\ref{fig:closure_DY_CR_on} and~\ref{fig:closure_Jets_CR_on} show representative closure-test results on selected observables from 
the Drell--Yan and dijet processes, respectively, for the \textbf{MPI-CRon} scenario; the variations of the four MPI parameters
are common to both the \textbf{MPI-CRoff} and \textbf{MPI-CRon} setups, while the $\eta_Q$ variation (orange) enters only the latter.
The largest impact on the observables is observed for the $\sigma_{\text{ND}}^{\rm norm}$ variations. Notably, for the charged-particle multiplicity
in Drell--Yan production the impact of variations of the CR parameter $\eta_Q$ is of similar strength as that of the matter-profile parameters.
We again remark that the $p_{\perp,\min}^{\rm ref}$ parameter can only be varied to larger values, i.e.\ here from $1.0\,\text{GeV}$ to $1.3\,\text{GeV}$.
In all variations the reweighted predictions are in very good agreement with the dedicated generator runs across the full distributions,
confirming that the reweighting procedure accurately captures the generator response to individual parameter variations over the full tuning ranges
considered.

\begin{figure}
\centering
    \includegraphics[width=1.0\textwidth]{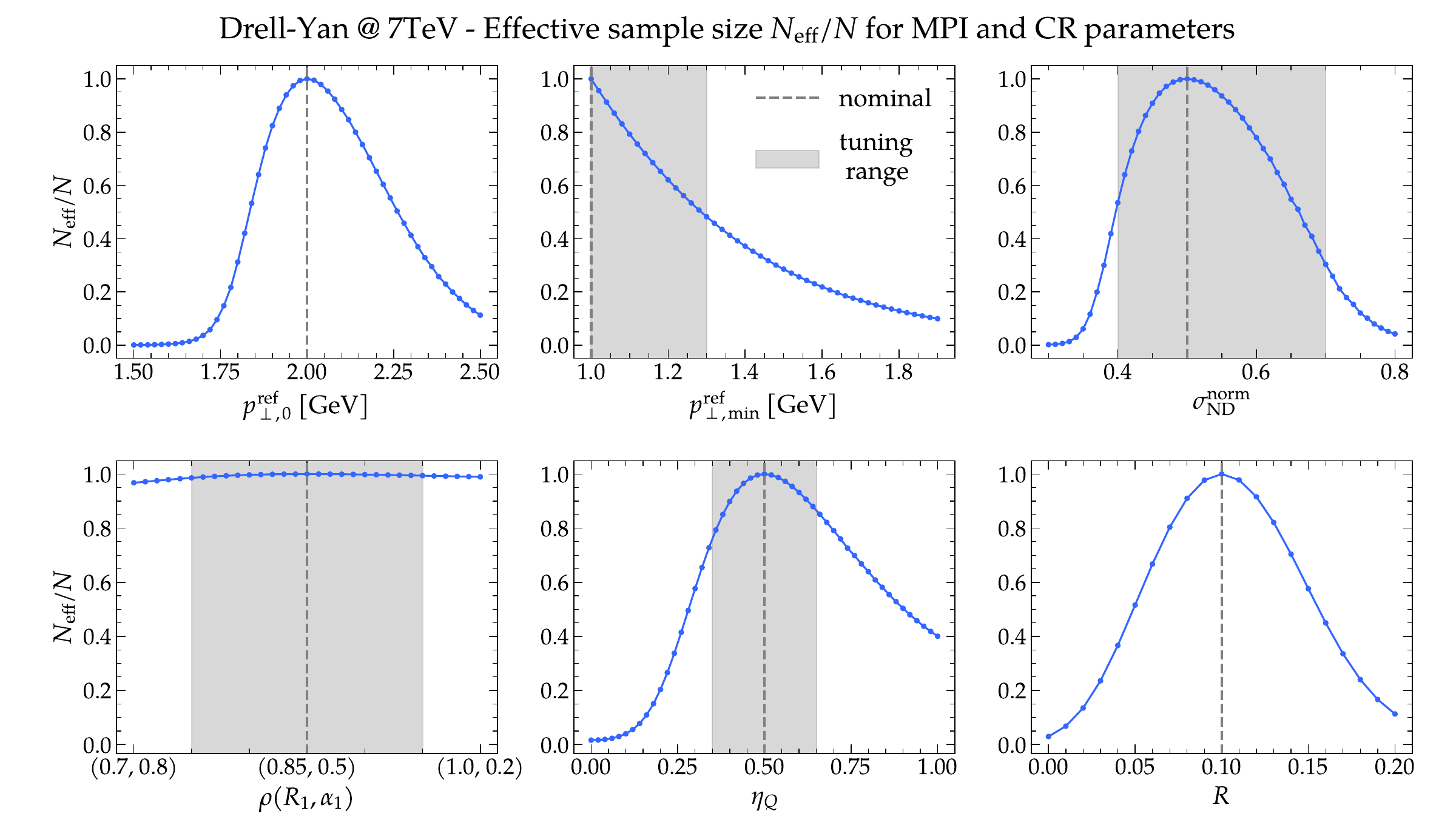}
\caption{The effective number of events after reweighting for single-parameter variations of the
MPI and CR models. Variations in $R_1$ and $\alpha_1$ are mapped to a joint variation in
$\rho(R_1, \alpha_1)$, which interpolates between the configurations of a dominant proton core
and a dominant halo. For each parameter, we display the ratio $N_{\rm eff}/N$ as a function of the
parameter value. Nominal values are indicated by vertical dashed lines. The shaded regions
visualise the parameter ranges used in tuning (cf.\ Table~\ref{tab:paramstunerange}).}
\label{fig:neff}
\end{figure}

To quantify the statistical dilution due to reweighting, introducing event-weight factors
$w^{(i)}$ for parameter variation $\vec{p}_i$, we consider the effective sample size

\begin{equation}
    N^{(i)}_{\rm eff} = \frac{\left(\sum_{j=1}^N w^{(i)}_j\right)^2}{\sum_{j=1}^N \left(w^{(i)}_j\right)^2}\,, 
    \label{eq:neff}
\end{equation}
where $N$ is the total number of nominal events generated for $\vec{p}_0$, while $w^{(0)}_j=1$ for all $j$
and thus $N^{(0)}_{\rm eff}=N$. Accordingly, this quantity measures the reduction in statistical power due to
broadly distributed or highly non-uniform event weights, and therefore provides a convenient measure
of the stability of the reweighting procedure. Figure~\ref{fig:neff} shows the effective sample size
relative to the nominal sample, i.e.\ $N^{(i)}_{\rm eff}/N$, for the same single-parameter variations used
in the closure tests above, including the combined $\rho(R_1, \alpha_1)$ variation.

Moving away from the nominal set the effective sample size drops quite quickly.
A notable exception are
the matter-profile parameters, which result in rather small weight variations. We here also indicate the
parameter ranges considered in the tuning. Also for the parameters inducing rather significant reweighting
factors, i.e.\ $p_{\perp,\min}^{\rm ref}$ and $\sigma_{\rm ND}^{\rm norm}$, even at the boundaries of the tuning
range the effective sample size is still ${\cal{O}}(N/2)$. These dilution effects need to be taken into account when
setting up the tuning process. First, sufficiently large event samples need to be considered such that
such a reduction in statistical power does not invalidate their usability for comparison with measurement
data of a given systematic and statistical uncertainty. Second, if the drop in the effective sample
size appears to be too large or wider parameter variations are considered, additional nominal samples
can be generated and reweighted accordingly. 

\subsection{Underlying Event tunes for \protect\Sherpa}\label{sec:tuneresults}

\begin{figure}
\centering
\begin{minipage}[t]{0.48\textwidth}
    \centering
    \includegraphics[width=1.0\textwidth]{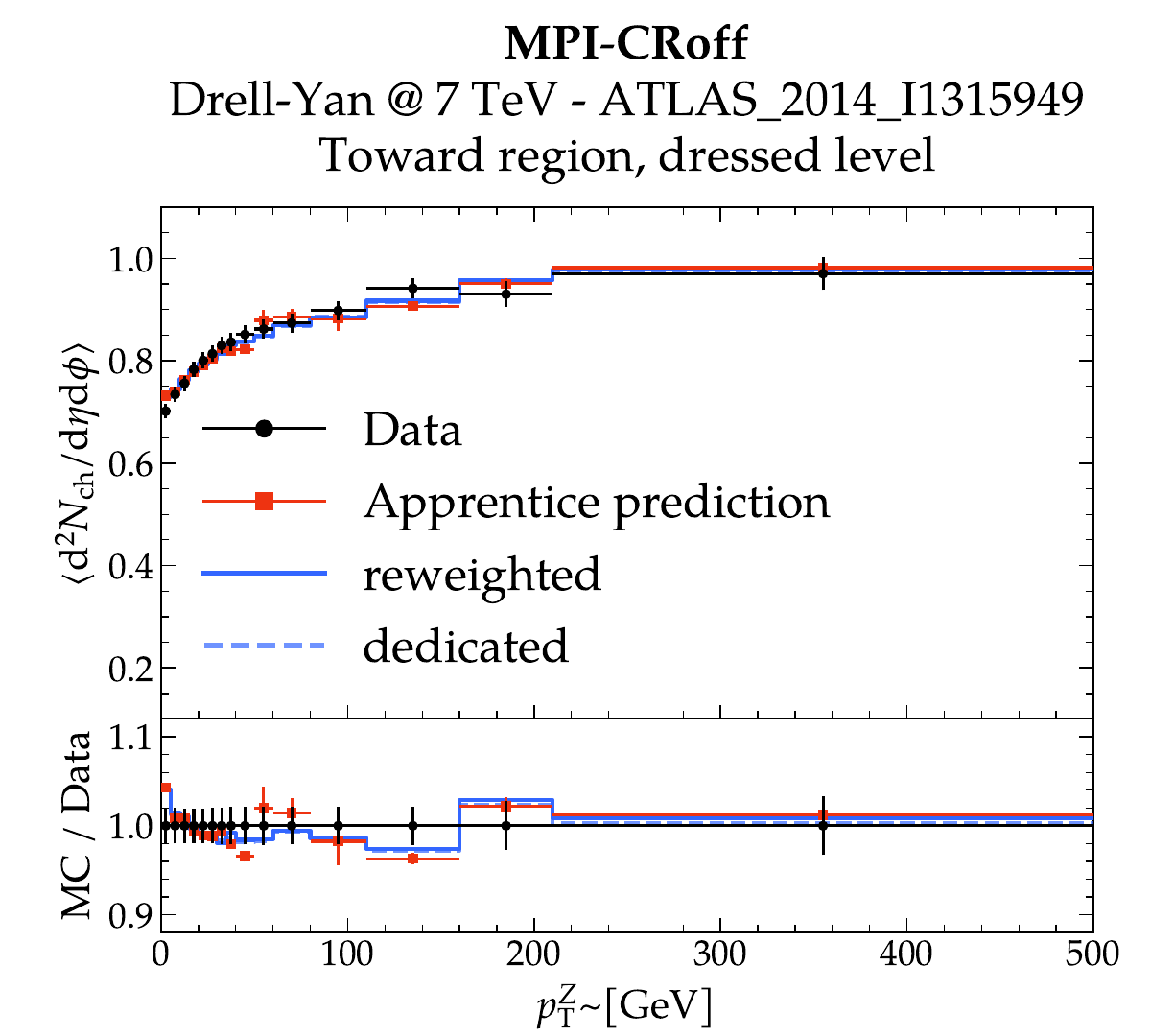}
    \caption{Results for the best-tune parameter set $\vec{p}^{\,*}$ for the \textbf{MPI-CRoff} scenario, compared to the ATLAS Drell--Yan underlying-event measurement~\cite{ATLAS:2014yqy}.}
    \label{fig:results_DY_CR_off}
\end{minipage}\hfill
\begin{minipage}[t]{0.48\textwidth}
    \centering
    \includegraphics[width=1.0\textwidth]{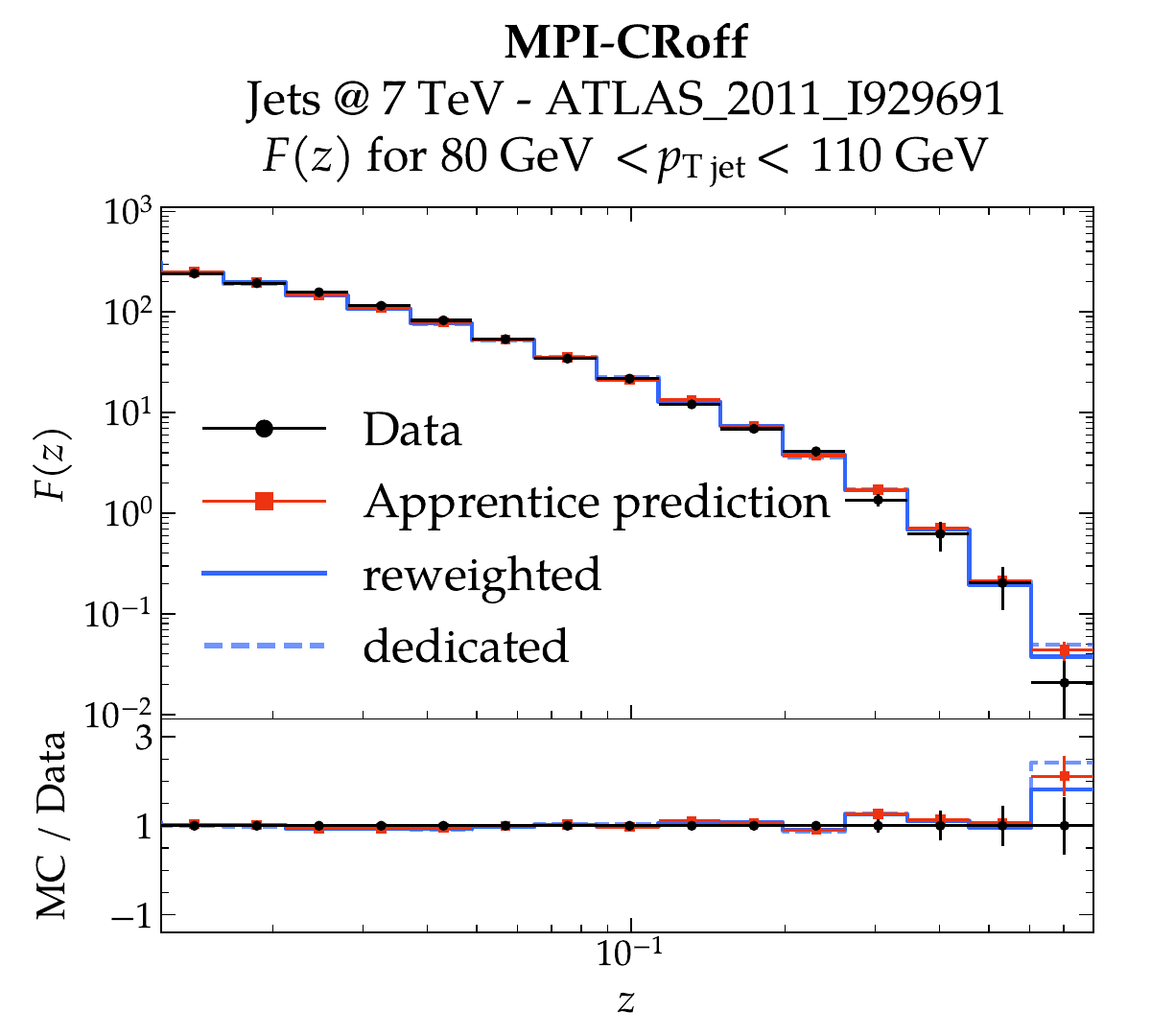}
    \caption{Results for the best-tune parameter set $\vec{p}^{\,*}$ for the \textbf{MPI-CRoff} scenario, compared to the ATLAS jet-fragmentation-function measurement~\cite{ATLAS:2011myc}.}
    \label{fig:results_Jets_CR_off}
\end{minipage}
\end{figure}

\begin{figure}
\centering
\begin{minipage}[t]{0.48\textwidth}
    \centering
    \includegraphics[width=1.0\textwidth]{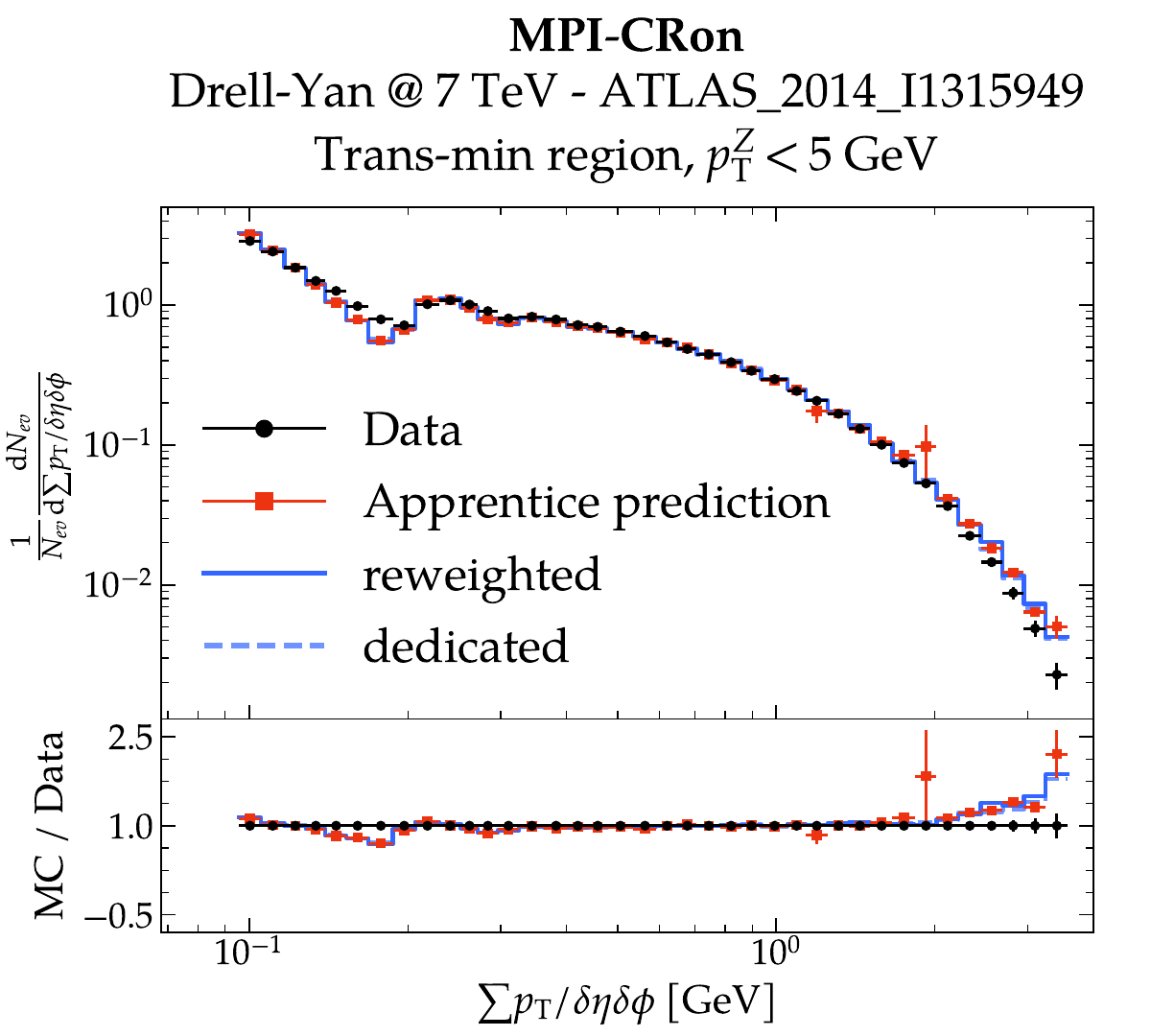}
    \caption{Results for the best-tune parameter set $\vec{p}^{\,*}$ for the \textbf{MPI-CRon} scenario, compared to the ATLAS Drell--Yan underlying-event measurement~\cite{ATLAS:2014yqy}.}
    \label{fig:results_DY_CR_on}
\end{minipage}\hfill
\begin{minipage}[t]{0.48\textwidth}
    \centering
    \includegraphics[width=1.0\textwidth]{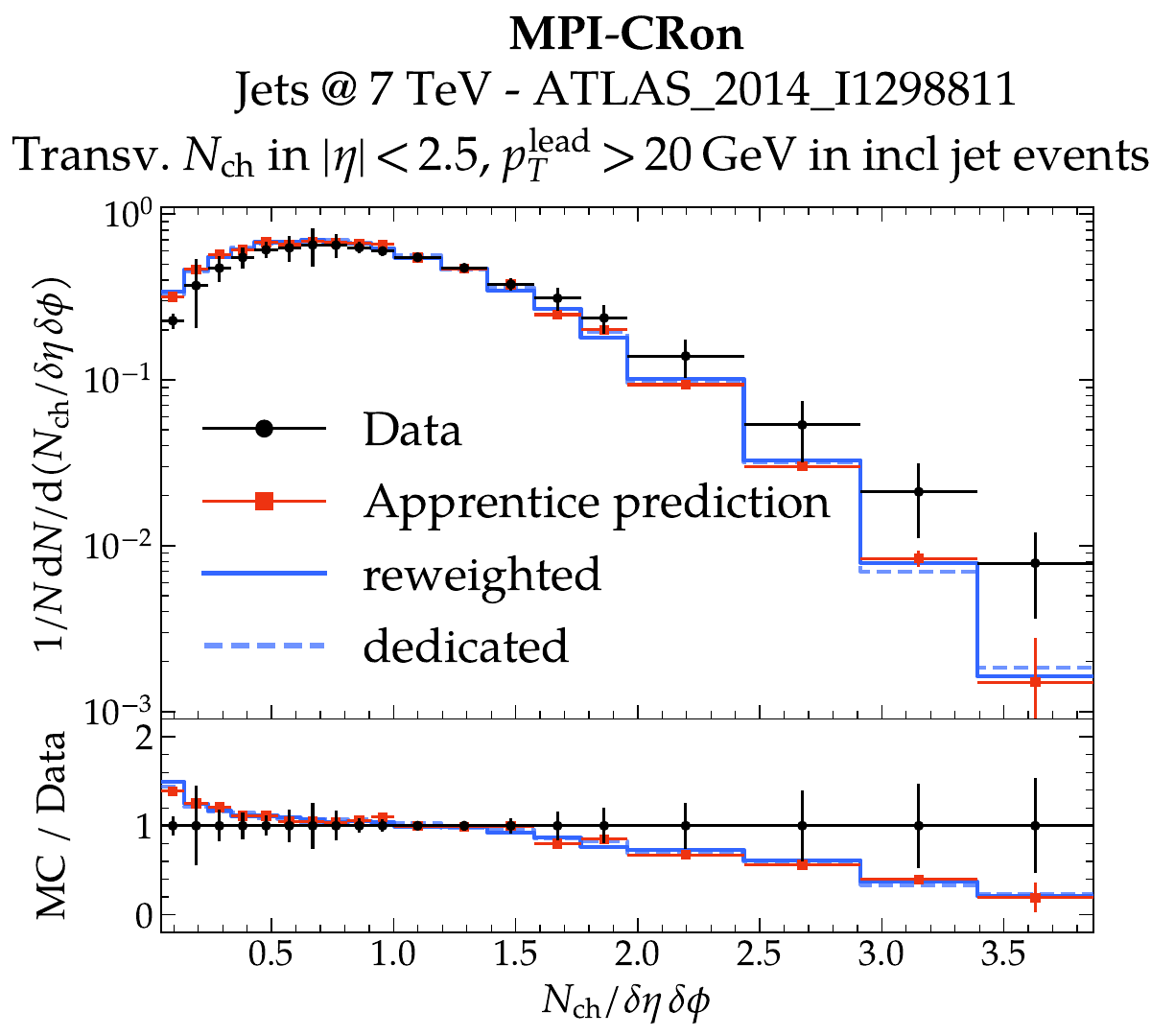}
    \caption{Results for the best-tune parameter set $\vec{p}^{\,*}$ for the \textbf{MPI-CRon} scenario, compared to the ATLAS underlying-event analysis in jet production~\cite{ATLAS:2014iez}.}
    \label{fig:results_Jets_CR_on}
\end{minipage}
\end{figure}

Having validated our novel reweighting method and with our corresponding tuning methodology
established we now move to the discussion of our resulting tunes of the \protect\Sherpa
MPI and CR models. We consider two situations, namely MPI with CR disabled and with CR
switched on. The best-tune parameter sets we obtain for the \textbf{MPI-CRoff} and
\textbf{MPI-CRon} scenarios are:
\begin{align}
  \text{\textbf{MPI-CRoff:}} \quad
  \vec{p}^{\,*} &= \left(R_1=0.93\,\text{fm},\ \alpha_1=0.32,\ \sigma_{\rm ND}^{\rm norm}=0.50,\
  p_{\perp,\min}^{\rm ref}=1.14\,\text{GeV}\right), \nonumber\\
  \text{\textbf{MPI-CRon:}} \quad
  \vec{p}^{\,*} &= \left(R_1=0.85\,\text{fm},\ \alpha_1=0.65,\ \sigma_{\rm ND}^{\rm norm}=0.44,\
  p_{\perp,\min}^{\rm ref}=1.10\,\text{GeV},\ \eta_Q=0.63\right). \nonumber
\end{align}
Figures~\ref{fig:results_DY_CR_off} and~\ref{fig:results_Jets_CR_off} show a representative selection of predictions at $\vec{p}^{\,*}$ 
for the \textbf{MPI-CRoff} tune, comparing to the ATLAS Drell--Yan underlying-event measurement~\cite{ATLAS:2014yqy} and the
ATLAS underlying-event analysis in jet production~\cite{ATLAS:2011myc}, respectively.
Selected \textbf{MPI-CRon} predictions are displayed in Figures~\ref{fig:results_DY_CR_on} and~\ref{fig:results_Jets_CR_on}.
In each figure we show two
theory predictions at $\vec{p}^{\,*}$: one obtained from a dedicated \Sherpa run with the best-tune parameters, and one obtained
by reweighting the nominal sample $\vec{p}_0$ to $\vec{p}^{\,*}$. The agreement between the two demonstrates that the reweighting
method is capable of faithfully reproducing simultaneous multi-parameter variations across all considered observables, without
the need for additional generator runs. The \textsc{Apprentice} surrogate prediction at $\vec{p}^{\,*}$ is overlaid,
showing that the employed polynomial approximation accurately captures the full generator response at the best-tune point.

The effective sample size ratios, cf.\ Eq.~\eqref{eq:neff}, and the average reweighting weights for all four runs
are summarised in Table~\ref{tab:ess_summary}. These quantities provide a compact measure of the statistical efficiency and
demonstrate that the weights remain well behaved, including for the combined MPI and CR variations. The full distributions of
the reweighting weights are shown in the Appendix~\ref{app2} in Figure~\ref{fig:total_weight_distribution}.

\begin{table}[h!]
\centering
  \caption{Summary of the effective sample size ratio $N_{\rm eff}/N$ and average reweighting weight $\langle w\rangle$.}
  \begin{tabular}{lll}
    \hline
    Sample & $N_{\rm eff}/N$ & $\langle w\rangle$\\
    \hline
    \textbf{MPI-CRoff} - Drell--Yan & $0.72$ & $1.000$ \\
    \textbf{MPI-CRon} - Drell--Yan & $0.58$ & $0.999$ \\
    \textbf{MPI-CRoff} - Jets & $0.72$ & $1.003$ \\
    \textbf{MPI-CRon} - Jets & $0.56$ & $0.994$ \\
    \hline
  \end{tabular}
  \label{tab:ess_summary}
\end{table}

\begin{table}
\centering
  \caption{$\chi^2$ values obtained for the best-tune parameter sets for the observables 
  considered in the \textbf{MPI-CRoff} and \textbf{MPI-CRon} tunes (Drell--Yan: 10\,M events, Jets: 15\,M events).\label{tab:chi2tuneresults}}
  \begin{tabular}{llll}
    \hline
    & \multicolumn{2}{c}{$\chi^2/{\rm ndf}$} & ndf\\
    Process/Analysis & \textbf{MPI-CRoff} & \textbf{MPI-CRon} &\\
    \hline
    Drell--Yan (cum.) & $34.0$ & $16.1$ & $2573$\\
    \hline
    \texttt{ATLAS\_2014\_I1315949} & $32.4$ & $14.7$ & $2123$\\
    \texttt{CMS\_2012\_I1107658} & $41.7$ & $22.6$ & $450$\\
    \hline
    Jets (cum.) & $2.5$ & $2.0$ & $1690$\\
    \hline
    \texttt{ATLAS\_2014\_I1298811} & $2.7$ & $2.1$ & $1035$\\
    \texttt{ATLAS\_2011\_I929691} & $1.6$ & $1.4$ & $409$\\
    \texttt{CMS\_2011\_I916908} & $3.6$ & $2.1$ & $163$\\
    \texttt{CMS\_2011\_I886332} & $3.1$ & $4.0$ & $78$\\
    \texttt{CMS\_2011\_I930319} & $1.1$ & $1.2$ & $5$\\
    \hline
  \end{tabular}
\end{table}

\begin{figure}
    \centering
    \includegraphics[width=1.0\textwidth]{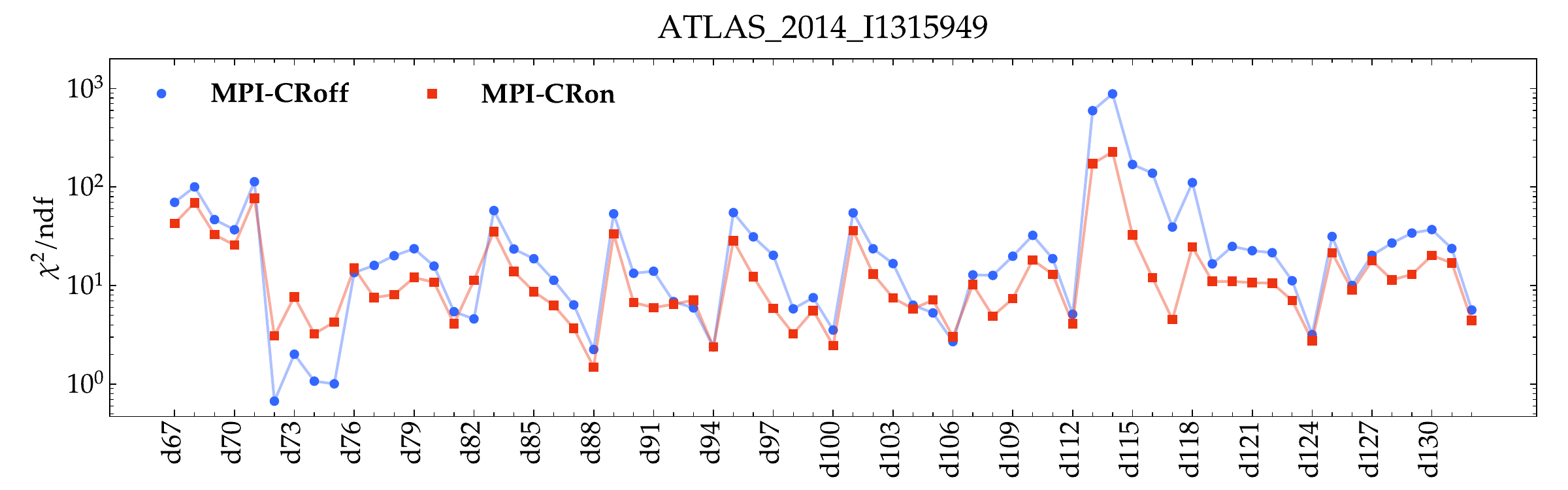}
    \includegraphics[width=1.0\textwidth]{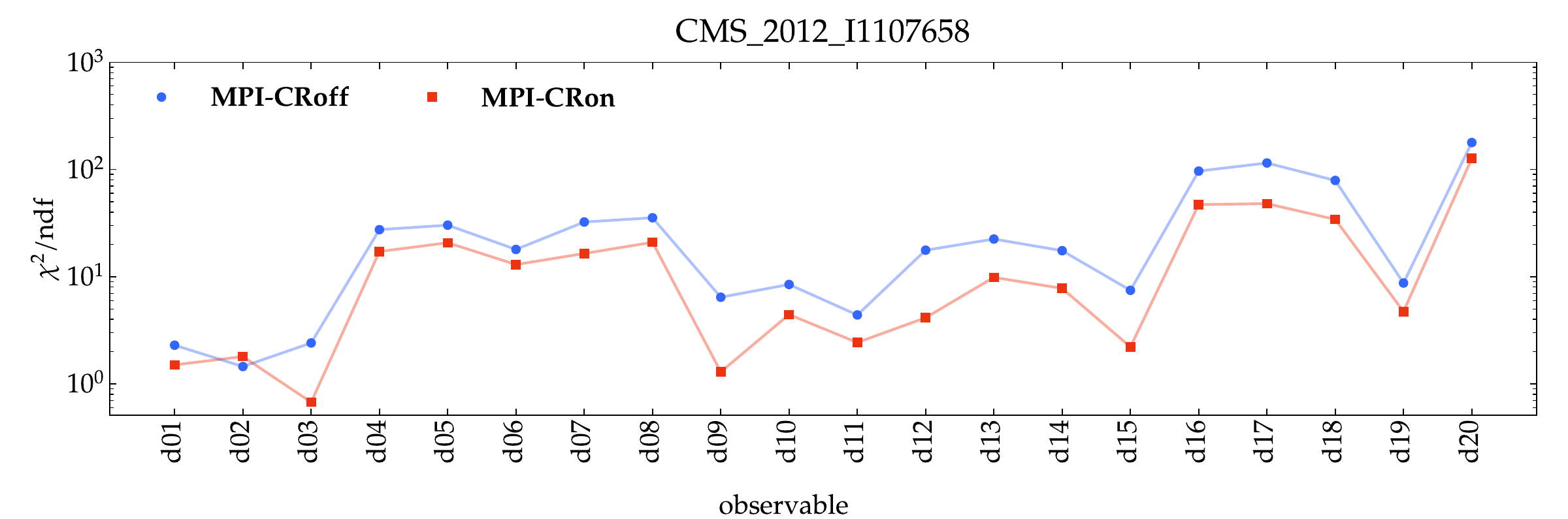}
    \caption{The $\chi^2/{\rm ndf}$ values per observable for all Drell--Yan analyses considered in the \textbf{MPI-CRoff} and \textbf{MPI-CRon} tunes.}
    \label{fig:chi2_plots_DY}
\end{figure}

A quantitative summary of the description quality at the best-tune points is provided in Table~\ref{tab:chi2tuneresults}, 
with per-observable $\chi^2/{\rm ndf}$ values shown in Figure~\ref{fig:chi2_plots_DY} for the two considered Drell--Yan analyses 
(analogous plots for the four jet-production analyses with more than one histogram appear in Appendix~\ref{app2}). 
The \textbf{MPI-CRon} tune yields a better overall agreement with the data than \textbf{MPI-CRoff}, both in the Drell--Yan and
jet sectors, but in particular in the former, where the $\chi^2/{\rm ndf}$ values are effectively halved. This establishes the
CR-enabled scenario as the preferred tune. The $\chi^2/\mathrm{ndf}$ values obtained here are consistent with those from a conventional
tune based on dedicated generator runs at each parameter point\footnote{A conventional tune was performed on a grid of dedicated
\Sherpa runs covering the same parameter ranges and observables, serving as an internal cross-check.}, confirming that the
reweighting-based approach is a reliable and computationally efficient alternative to a tune based on dedicated simulation runs.

\begin{figure}
\centering
\begin{minipage}[t]{0.48\textwidth}
    \centering
    \includegraphics[width=1.0\textwidth]{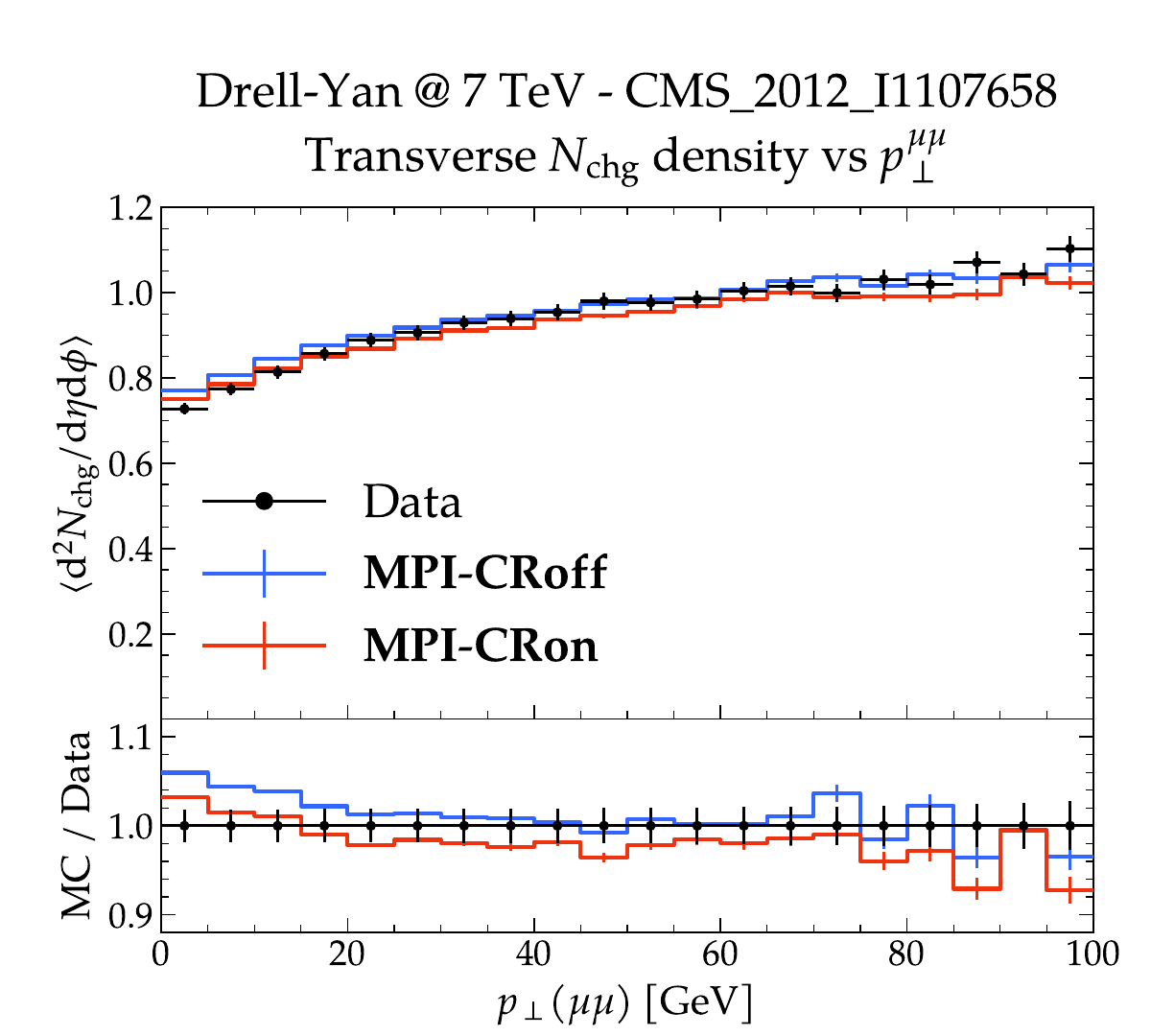}
    \includegraphics[width=1.0\textwidth]{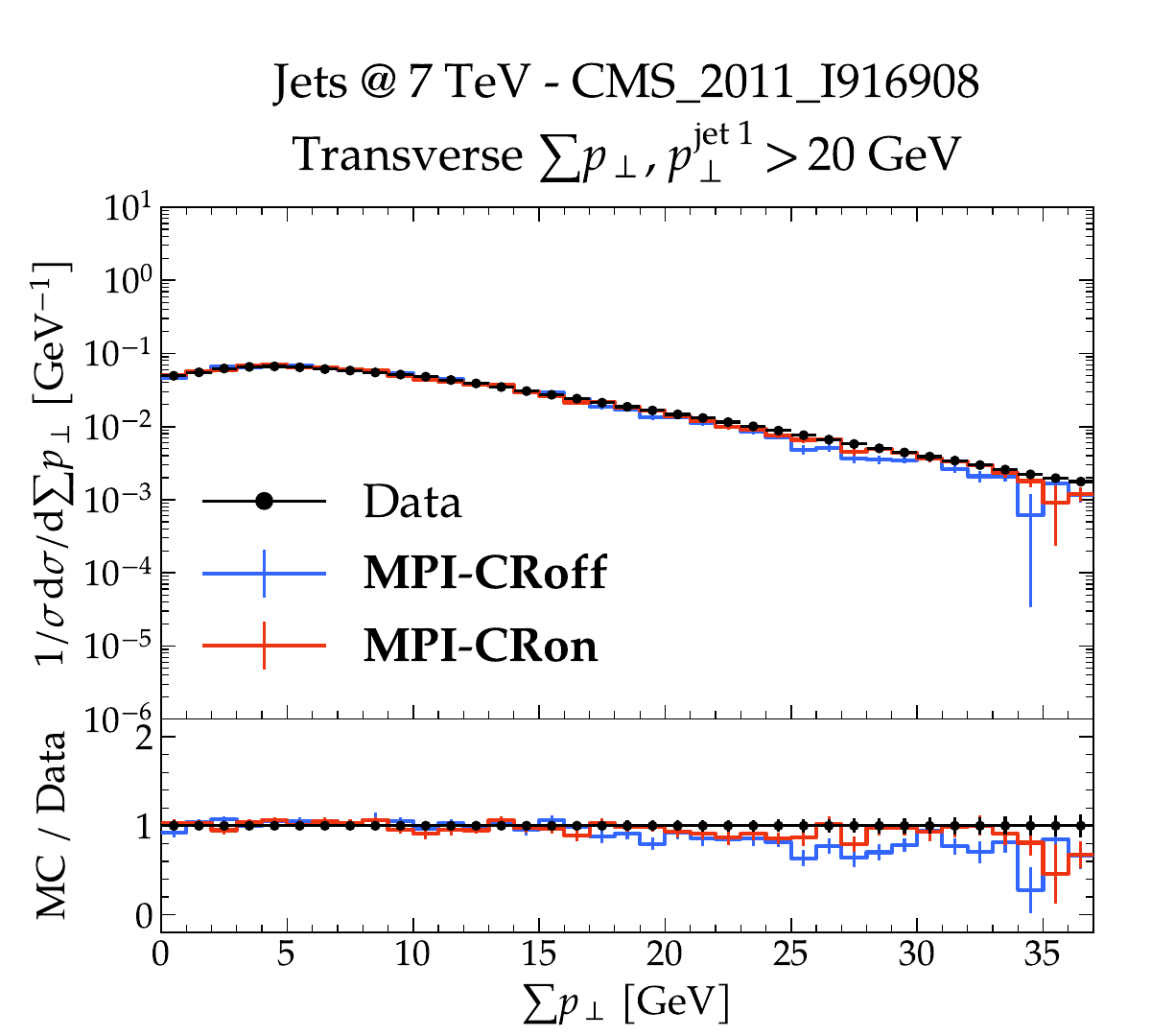}
\end{minipage}\hfill
\begin{minipage}[t]{0.48\textwidth}
    \centering
    \includegraphics[width=1.0\textwidth]{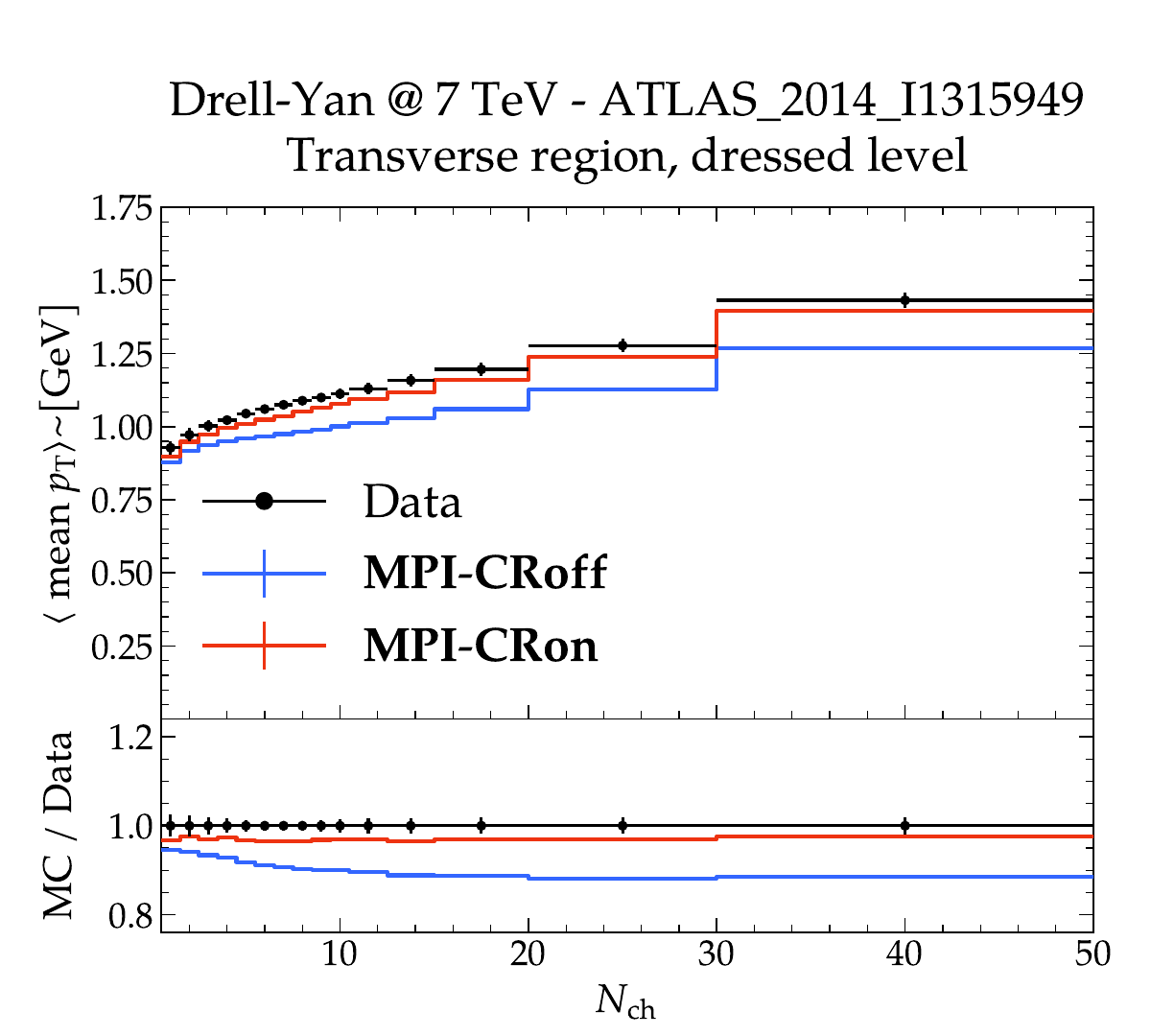}
    \includegraphics[width=1.0\textwidth]{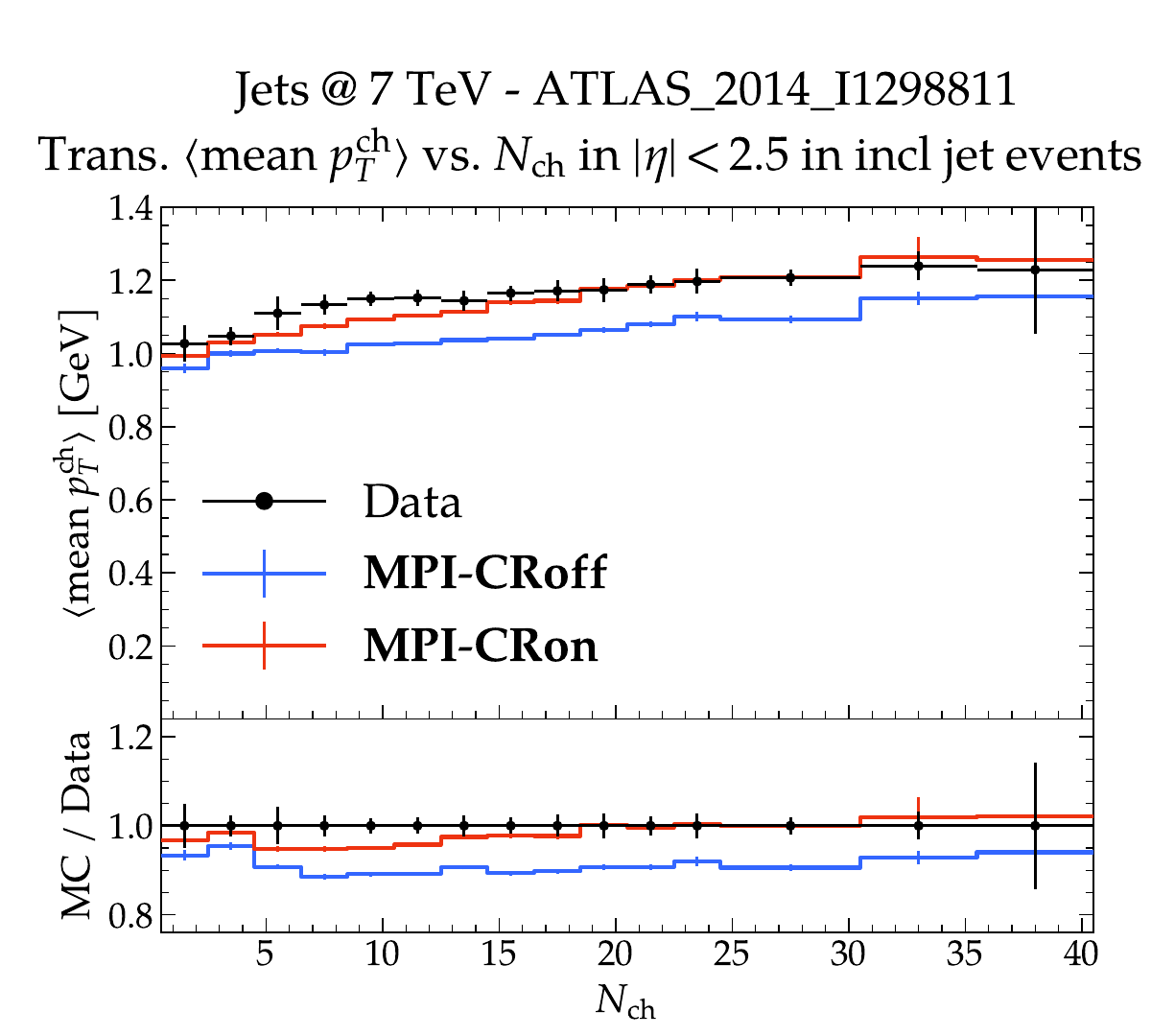}
\end{minipage}
\caption{Direct comparison of the \textbf{MPI-CRoff} (blue) and \textbf{MPI-CRon} (red) best-tune predictions with data, for representative underlying-event observables in
Drell--Yan (top) and jet (bottom) production at $\sqrt{s}=7\,\text{TeV}$. Left: the transverse-region charged-particle density versus the di-muon transverse momentum~\cite{CMS:2012oqb}
and the transverse-region charged $p_\perp$ sum~\cite{CMS:2011qzf}. Right: the mean charged-particle transverse momentum as a function of multiplicity in the transverse region,
from the ATLAS Drell--Yan~\cite{ATLAS:2014yqy} and ATLAS jet~\cite{ATLAS:2014iez} measurements.}
\label{fig:results_comp}
\end{figure}

To make the impact of colour reconnection on the physical distributions explicit, Figure~\ref{fig:results_comp} directly compares
the \textbf{MPI-CRoff} and \textbf{MPI-CRon} best-tune predictions with data for a representative set of Drell--Yan and jet observables
used in the tune. Colour reconnection, which rearranges the colour flow prior to cluster formation and thereby
increases the average transverse momentum of the produced hadrons at a given multiplicity, improves the description of
observables that correlate $\langle p_\perp\rangle$ with the charged-particle multiplicity (right column), which is largely driven
by the underlying event. Here the tune without CR undershoots the data and yields too weak a rise of $\langle p_\perp\rangle$ with
$N_{\rm ch}$, whereas the \textbf{MPI-CRon} tune recovers the measured trend without degrading the already good description of
the observables in the left column. This improvement is the main source of the lower $\chi^2/{\rm ndf}$ obtained for CR-enabled-scenario
in Table~\ref{tab:chi2tuneresults}.

\subsection{Energy extrapolation}
\label{sec:energyextrapolation}

\begin{table}[t!]
\centering
  \caption{$\chi^2$ values obtained for the best-tune energy-extrapolation parameter $\eta^*$ for the observables considered in the tune (15\,M events each).\label{tab:chi2tuneeta}}
  \begin{tabular}{lll}
    \hline
    & \multicolumn{1}{c}{$\chi^2/{\rm ndf}$} & ndf\\
    Analysis & \textbf{MPI-CRon} &\\
    \hline
    \texttt{ATLAS\_2019\_I1736531} & $28.6$ & $14927$\\
    \texttt{CDF\_2015\_I1388868} & $16.3$ & $120$\\
    \hline
  \end{tabular}
\end{table}

Having established best-tune parameter sets at a reference scale $s_\text{ref}$, here $\sqrt{s_\text{ref}}=7\,\text{TeV}$,
we still need to determine the scaling parameter $\eta$ that according to Eq.~\eqref{eq:pts} determines the values of the
dimensionful MPI parameters at an alternative scale. To this end, we perform a dedicated one-parameter \textsc{Apprentice}
tune of $\eta$ using underlying-event measurements at two different centre-of-mass energies: an ATLAS Drell--Yan underlying-event 
analysis at $\sqrt{s}=13\,\text{TeV}$~\cite{ATLAS:2019ocl} (\texttt{ATLAS\_2019\_I1736531}) and a CDF track-based underlying-event
study in $p\bar{p}$ collisions at $\sqrt{s}=1.96\,\text{TeV}$~\cite{CDF:2015txs} (\texttt{CDF\_2015\_I1388868}). 

In both cases we employ the \textbf{MPI-CRon} tune from the previous section as a baseline and use the reweighting procedure to
scan $\eta$ in the range $[0.0, 0.08]$, while keeping all other parameters fixed 
at their best-tune values. Note, variations of $\eta$ are constrained by the same phase-space limitation as $p_{\perp,\min}^{\rm ref}$,
i.e.\ only variations in one direction are possible. The direction thereby depends on whether the target collision energy $\sqrt{s}$
exceeds or falls below the reference scale. Consequently, for the $13\;\text{TeV}$ LHC runs we use $\eta=0$ as nominal value, for
the Tevatron setup instead $\eta=0.08$ is used as baseline.

\begin{figure}
    \centering
    \includegraphics[width=1.0\textwidth]{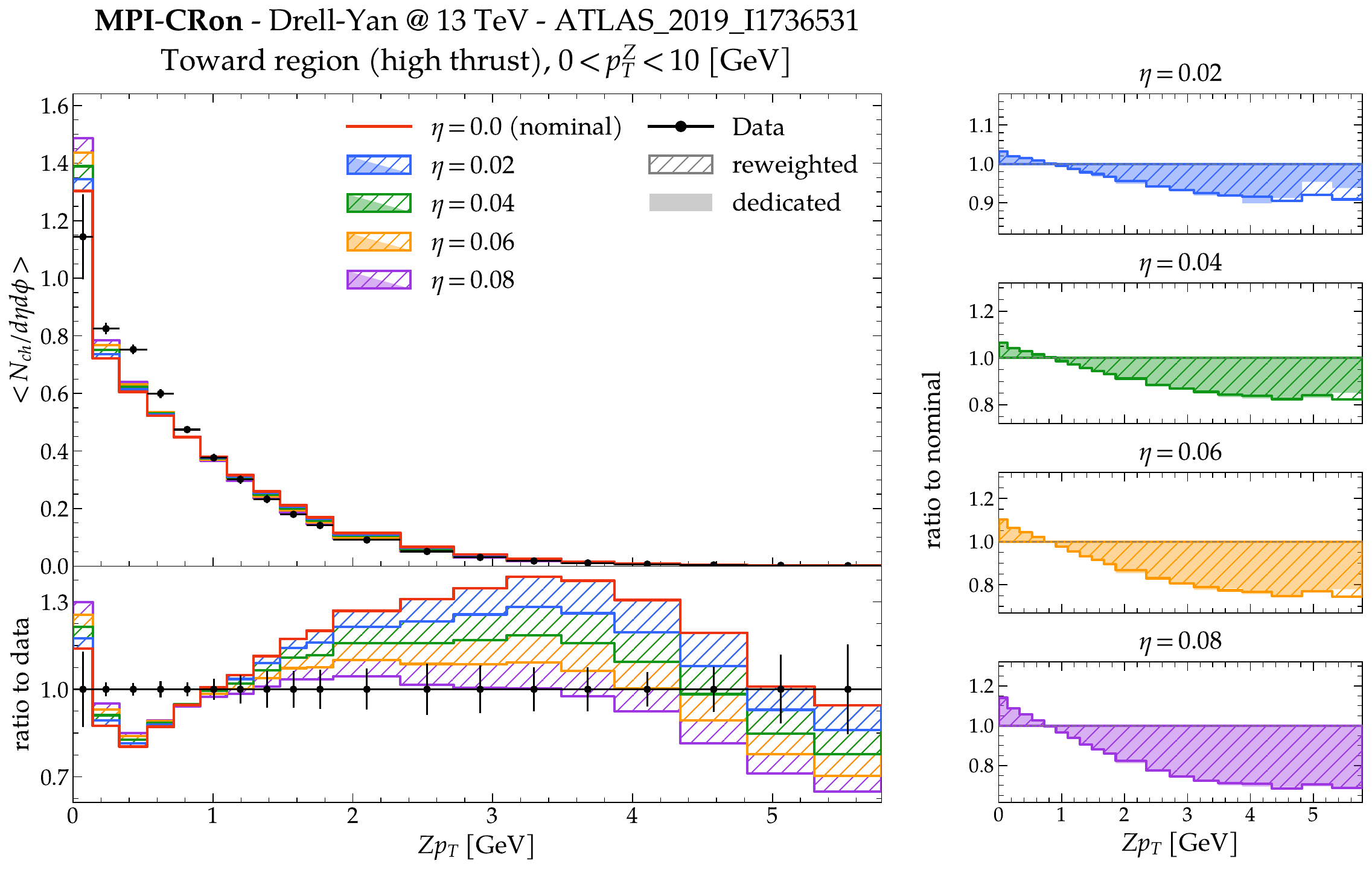}
    \caption{Closure test for the MPI energy-extrapolation parameter $\eta$ using the ATLAS underlying-event measurement in inclusive $Z$-boson production at $\sqrt{s}=13\,\text{TeV}$~\cite{ATLAS:2019ocl}.
    The left panel shows the variations obtained from the reweighting procedure of $\eta$ for the \textbf{MPI-CRon} best-tune parameter set. The panels on the right compare the reweighting-based variations 
    to those obtained from dedicated generator runs for the corresponding $\eta$ values.
    }
    \label{fig:closure_eta_DY}
\end{figure}

\begin{figure}
    \centering
    \includegraphics[width=1.0\textwidth]{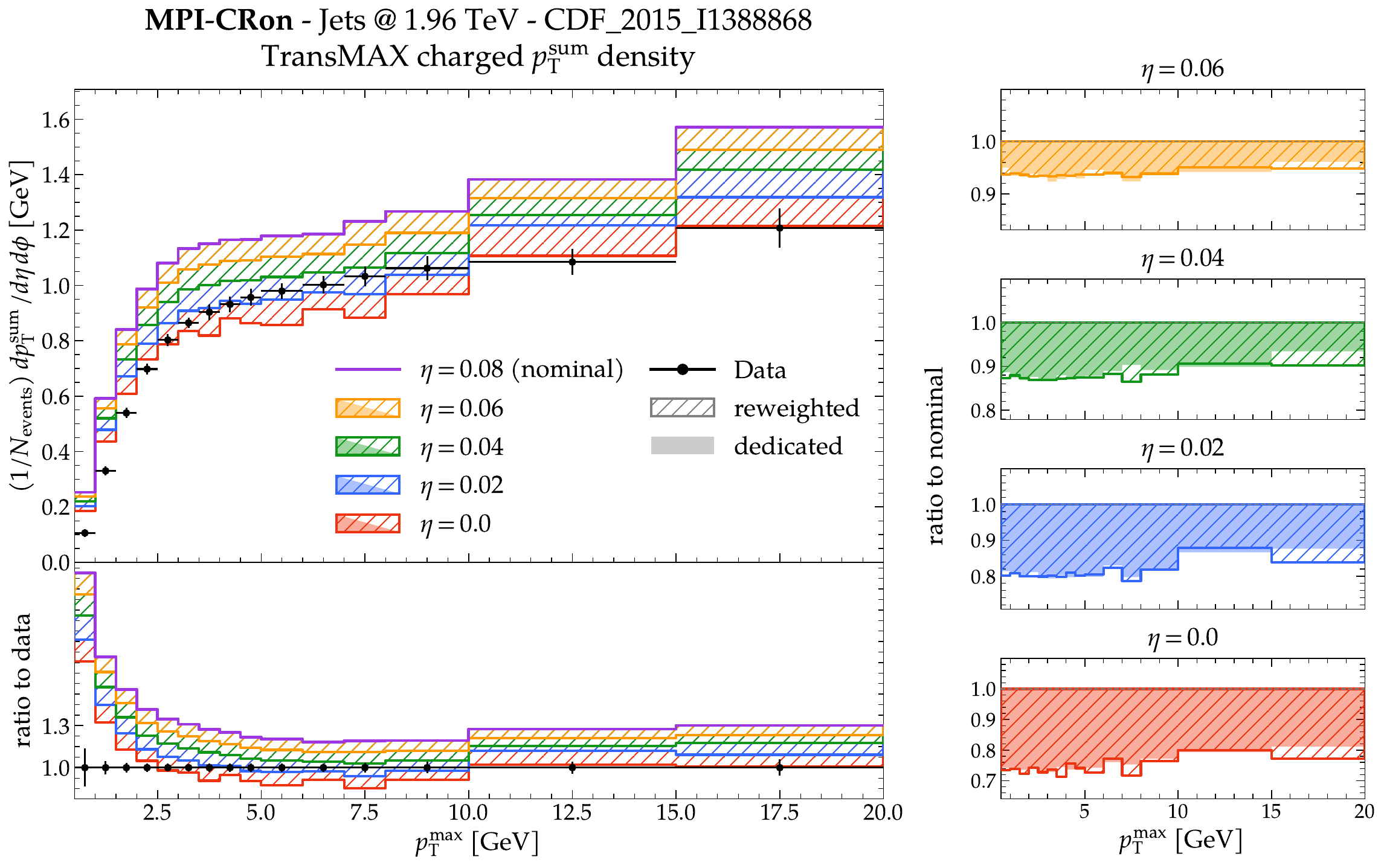}
    \caption{Closure test for the MPI energy-extrapolation parameter $\eta$ using the CDF underlying-event measurement in $p\bar{p}$ collisions at $\sqrt{s}=1.96\,\text{TeV}$~\cite{CDF:2015txs}.
    The left panel shows the variations obtained from the reweighting procedure of $\eta$ for the \textbf{MPI-CRon} best-tune parameter set. The panels on the right compare the reweighting-based variations to those obtained from dedicated generator runs for the corresponding $\eta$ values.
    }
    \label{fig:closure_eta_Jets}
\end{figure}

\begin{figure}
\centering
\begin{minipage}[t]{0.49\textwidth}
    \centering
    \includegraphics[width=1.0\textwidth]{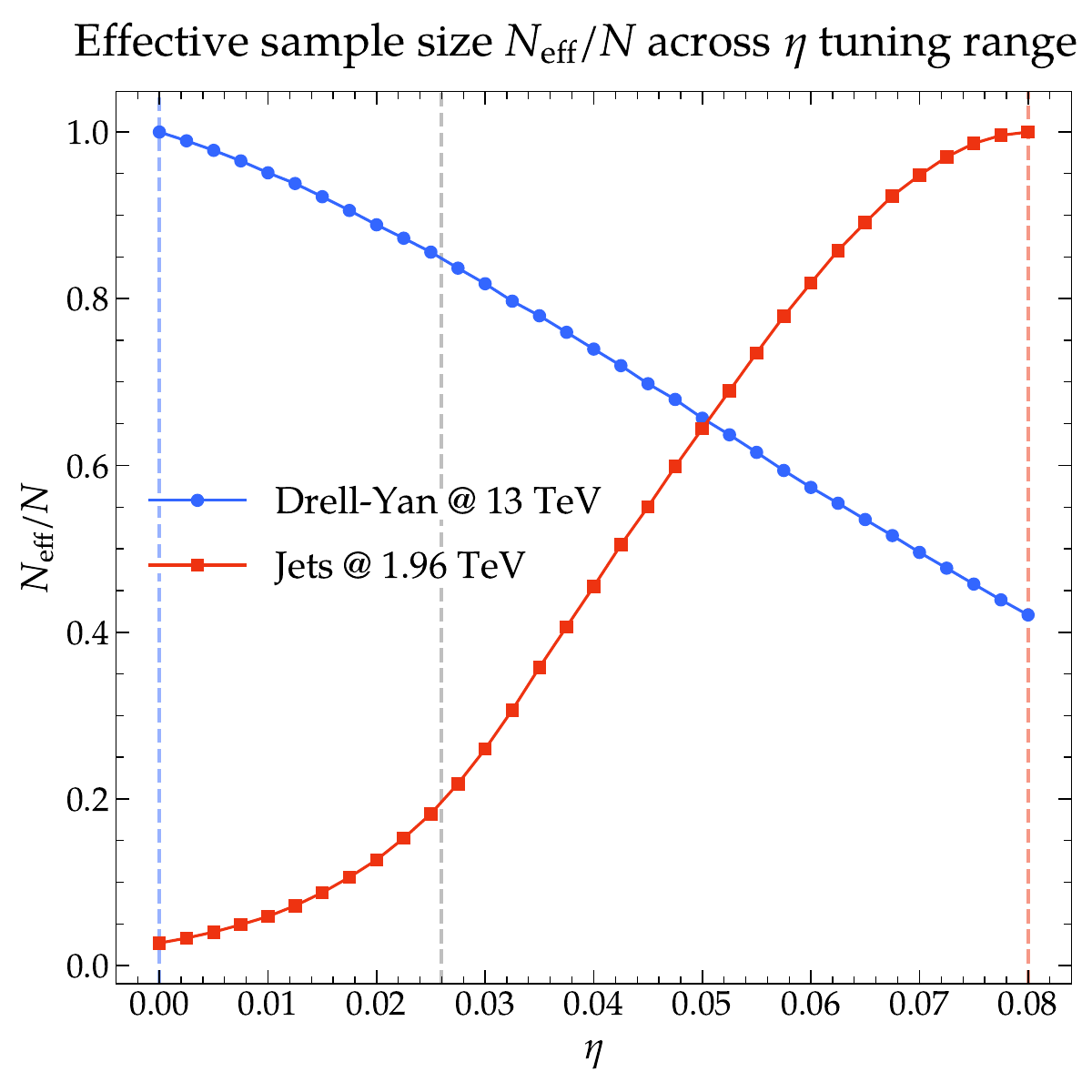}
\end{minipage}\hfill
\caption{The effective sample size $N_{\rm eff}/N$ as a function of the energy-extrapolation exponent $\eta$ across the tuning range employed in Section~\ref{sec:energyextrapolation}, 
for the Drell--Yan sample at $\sqrt{s}=13\,\text{TeV}$ and the jet sample at $\sqrt{s}=1.96\,\text{TeV}$. The nominal values, $\eta=0.0$ (for $13\,\text{TeV}$) and $\eta=0.08$ 
(for $1.96\,\text{TeV}$), are indicated by coloured dashed lines. Reweighting is performed only in the direction permitted by the phase-space constraint. The tuned value, 
$\eta^*=0.026$, is shown by the grey dashed line.}
\label{fig:neff_eta_plot}
\end{figure}

For the tuning grid we consider $18$ uniformly spaced $\eta$ values, and the surrogate model is
constructed as a simple polynomial function of degree three. A representative validation of the energy extrapolation across
the entire tuning range is shown in Figures~\ref{fig:closure_eta_DY} and~\ref{fig:closure_eta_Jets} for the LHC and the Tevatron
analysis, respectively. In both cases the reweighted predictions for different values of $\eta$ are in very good agreement
with the dedicated \Sherpa runs across the full distributions. The effective sample size for the reweighted samples is shown in
Figure~\ref{fig:neff_eta_plot}.

The two reference measurements are combined following the same prescription as for the 7\,TeV tune (cf.\ Section~\ref{sec:tuning_method}): 
the per-observable weights of the ATLAS (13\,TeV) and CDF (1.96\,TeV) data sets are rescaled such that the extracted value of $\eta$ is 
not driven by a single energy point. From the combined description we obtain an optimal value of $\eta^* = 0.026$ for the energy-extrapolation 
exponent. The $\chi^2/{\rm ndf}$ values for both analyses at $\eta^*$ are shown in Table~\ref{tab:chi2tuneeta}.

\begin{figure}
\centering
\begin{minipage}[t]{0.48\textwidth}
    \centering
    \includegraphics[width=1.0\textwidth]{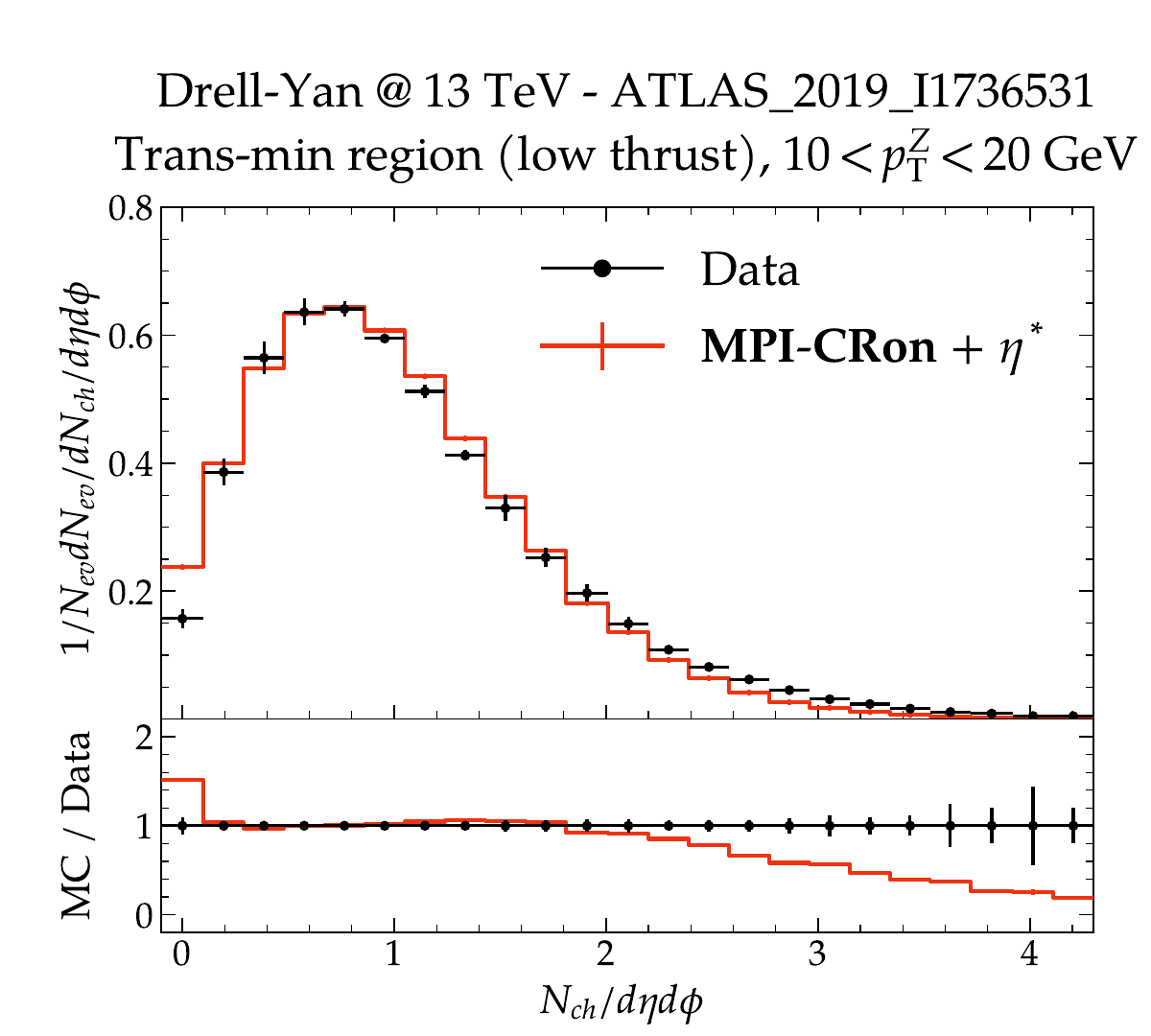}
\end{minipage}\hfill
\begin{minipage}[t]{0.48\textwidth}
    \centering
    \includegraphics[width=1.0\textwidth]{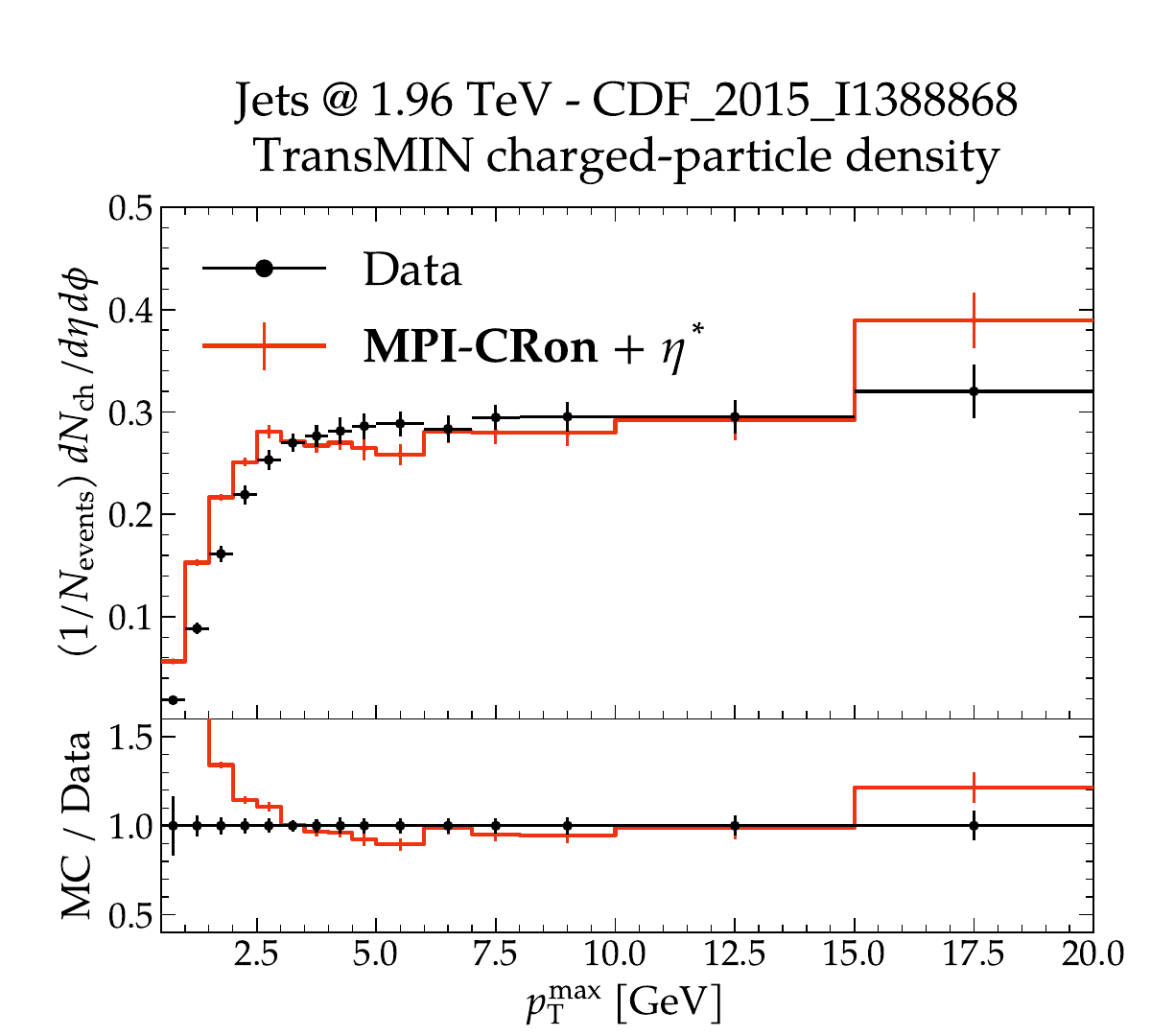}
\end{minipage}
\caption{Best-tune \textbf{MPI-CRon} predictions supplemented with the energy-extrapolation exponent $\eta^*=0.026$ (red), compared with data. 
Left: the trans-min charged-particle density in the low-thrust region for $10<p_T^Z<20\,\text{GeV}$, from the ATLAS inclusive
$Z$-boson underlying-event measurement at $\sqrt{s}=13\,\text{TeV}$~\cite{ATLAS:2019ocl}. Right: the trans-min charged-particle density as a function of the 
leading-particle transverse momentum $p_T^{\rm max}$, from the CDF underlying-event measurement in proton--anti-proton
collisions at $\sqrt{s}=1.96\,\text{TeV}$~\cite{CDF:2015txs}.}
\label{fig:results_eta}
\end{figure}

We close the discussion of the energy-scaling of MPI parameters by showing two exemplary observable distributions for the
\textbf{MPI-CRon} best-tune parameter set supplemented by $\eta^*=0.026$ in Fig.~\ref{fig:results_eta}. In the left panel
we show the differential distribution of the charged-particle density in the trans-min region with respect to the $Z$ boson,
comparing to data from the ATLAS experiment~\cite{ATLAS:2019ocl}. The right panel, instead, shows for the same region, but
defined with respect to the leading charged particle, the charged-particle density differential in the leading charged particle
transverse momentum, comparing to data from CDF~\cite{CDF:2015txs}. For both distributions \Sherpa provides a good description
of the data, although there is certainly space for improvement. Furthermore, with the advent of the high-luminosity LHC and
more analyses at $13$ and $13.6\,\text{TeV}$ becoming available, a dedicated tune at these higher energies, without
having to rely on the energy-scaling hypothesis, is well motivated. 

\section{Conclusions and Outlook}\label{sec:conclusions}

In this paper we have presented the first implementation and careful validation of an on-the-fly reweighting
algorithm to account for model-parameter variations in MPI and CR simulations with the \Sherpa
event-generator framework. In a single run, using a nominal set of parameters, our method 
evaluates in-situ event-wise variational weights corresponding to alternative points
in model-parameter space. We have performed detailed closure tests, both for model-internal variables
and actual physical observables, proving that the corresponding predictions agree with the outcome of
dedicated simulator runs. While in principle parameters can be varied largely at will, there are inherent
limitations. Certain parameters, whose settings restrict the phase space, can only be varied unidirectionally.
For the case of MPI this applies in particular to the transverse-momentum cutoff
of the secondary scatterings, i.e.\ $p_{\perp,\min}^{\rm ref}$, which, with respect to the nominal value, can
only be reweighted for upward variations. The introduction of additional event weights leads to a 
statistical dilution of the variational samples with respect to the nominal case, an effect that can be
measured by the Kish effective sample size. The reduced statistical power for parameter points significantly
away from the nominal set can potentially limit the usability of the corresponding predictions. This
limitation, however, can easily be counteracted by the introduction of additional nominal runs, and, as
far as computationally feasible, by increasing the statistics of the nominal sample.

The benefits of the reweighting technique are manifold: based on a single (or potentially a few) simulator
runs, it allows us to produce a significant number of alternative predictions with rather little
computational overhead, paying off for two variations already. The computational needs for
sample production are massively reduced. The same holds for storage and post-processing, e.g.\ in an analysis or when
invoking a detector simulation, where only a single sample, supplemented by alternative event weights, needs
to be processed. This qualifies our method for example for parameter-sensitivity studies, model-calibration
and tuning tasks, as well as uncertainty quantification at drastically reduced computational cost compared
to the current standard procedures.

As a first illustrative use case we have applied the method in the combined tuning of the \Sherpa MPI
and CR models. To this end, we considered underlying-event analyses in proton--proton collisions
at $\sqrt{s}=7\,\text{TeV}$, and produced corresponding generator predictions for $500$ random samples
from the $4$-dimensional (\textbf{MPI-CRoff}) and $5$-dimensional (\textbf{MPI-CRon}) parameter spaces
via reweighting from a single nominal run, respectively. These predictions are then used to train
polynomial surrogate models using the \textsc{Apprentice} package. Through comparison with data we
find best-tune parameter sets that provide an improved description of the considered Drell--Yan and
jet-production data. We find that the experimental data favours the inclusion of colour reconnections
in the simulation. 

To fix the scaling of dimensionful model parameters with the collision energy we furthermore performed
a simple one-parameter tune of the relevant scaling exponent, using LHC $pp$ data at $13\,\text{TeV}$
and $p\bar{p}$ data at $1.96\,\text{TeV}$ from the Tevatron. For both energies, the corresponding
predictions for $18$ parameter points have been obtained via reweighting from a nominal value, respectively.
With the obtained optimal parameter, the rescaled versions of our best-tune \textbf{MPI-CRon} parameter
set extracted at $\sqrt{s}=7\,\text{TeV}$ provide a satisfactory description of the considered data.

Our reweighting method for MPI and CR simulations closes a significant gap in the capabilities of modern
Monte Carlo event generators to efficiently provide predictions for alternative non-perturbative model
parameters. While parameter calibration, i.e.\ generator tuning, is an attractive application of such
a reweighting technique, we envisage that in practice it has the largest impact on parameter-sensitivity
studies and, in particular, the quantification of non-perturbative uncertainties. In particular for the
recently presented History Matching calibration method~\cite{Iskauskas:2026rxi} reweighting is a natural
fit, as samples from a non-implausible parameter space can be used to reflect the inherent parametric
uncertainties of a complex model. To this end, we plan to transfer our approach to \Sherpa's cluster-fragmentation
model~\cite{Chahal:2022rid,Knobbe:2025cef}, as well as minimum-bias simulations, and the generation of
the intrinsic transverse momenta. Ultimately, we aim to provide support for the reweighting of all
non-perturbative parameters in \Sherpa. 
The method presented here will pave the way to a new approach to robustly and systematically
assess parametric uncertainties of non-perturbative models, and will be essential for
the upcoming high-luminosity era of \textsc{LHC} operations.

\section*{Acknowledgements}
S.S.\ is grateful for financial support from the German Federal Ministry of Research, Technology and Space (projects 05D23MG1 and 05H24MGA).
This material is based upon work supported by Fermi Forward Discovery Group, LLC under Contract No. 89243024CSC000002 with the U.S. Department of Energy, Office of Science, Office of High Energy Physics. The work of M.K.\ was supported by the U.S. Department of Energy, Office of Science, Office of Advanced Scientific Computing Research, Scientific Discovery through Advanced Computing (SciDAC-5) program, grant “NeuCol”.
F.K.\ acknowledges funding by STFC under grant agreement ST/P006744/1.

\appendix

\section{Configuring MPI and CR Variations}\label{app}

The reweighting functionality for the \Sherpa MPI and CR models can be enabled by specifying lists of parameter values
 directly in the run card\footnote{As of version \Sherpa-3.0 these follow the \textsc{Yaml}~\cite{YAML2009} syntax.}, 
 instead of single values. The first element in each list corresponds to the nominal value (used for the main \Sherpa event generation), 
 while the remaining entries define the parameter variations for which additional weights will be computed.

    As an example, the run card setting 
    \begin{verbatim}
    SIGMA_ND_NORM: [0.5, 0.4, 0.6]\end{verbatim}
    will use $0.5$ as nominal value for the $\sigma_{\rm ND}^{\rm norm}$ parameter and additionally compute reweighted event weights
    corresponding to the alternative values $0.4$ and $0.6$.

    When multiple parameters are varied simultaneously, the variations are matched by their position in the respective lists. 
    That is, the first variation entry of each parameter forms one variation set, the second entries forms the next set, and so on. For example,
    \begin{verbatim}
    SIGMA_ND_NORM: [0.5, 0.4, 0.6]
    PT_Min(ref): [1.0, 1.2, 1.4]\end{verbatim}
    will produce a \Sherpa run with the nominal parameter set
    $$\vec{p}_0=(\texttt{SIGMA\_ND\_NORM},\texttt{PT\_Min(ref)})=(0.5,1.0)\,,$$
    and generate variation weights for
    $$\vec{p}_1=(0.4,1.2),\quad\vec{p}_2=(0.6,1.4)\,.$$
    If the number of variation entries differs between parameters, missing values are automatically filled with the corresponding nominal value. For instance,
    \begin{verbatim}
    SIGMA_ND_NORM: [0.5, 0.4, 0.6]
    PT_Min(ref): [1.0, 1.2]\end{verbatim}
    results in the two variations
    $$\vec{p}_1=(0.4,1.2),\quad\vec{p}_2=(0.6,1.0)\,.$$
    The same syntax applies to all parameters controlling the MPI and CR models listed in Table~\ref{tab:MPIparams} and Table~\ref{tab:CRparams}. 
    Mixed variations of both parameter groups are also supported, and \Sherpa will combine the respective weights accordingly. For example
    \begin{verbatim}
    AMISIC:
      SIGMA_ND_NORM: [0.5, 0.4, 0.6]

    COLOUR_RECONNECTIONS:
      ETA_Q: [0.5, 0.4, 0.6]\end{verbatim}
    will produce a \Sherpa run with the nominal parameter set
    $$\vec{p}_0=(\texttt{SIGMA\_ND\_NORM},\texttt{ETA\_Q})=(0.5,0.5)\,,$$
    and the variations
    $$\vec{p}_1=(0.4,0.4),\quad\vec{p}_2=(0.6,0.6)\,.$$
    The reweighting outputs, i.e.\ the variational event weights, are available both in the \textsc{HepMC}~\cite{Buckley:2019xhk} event record 
    and through the \Sherpa internal \textsc{Rivet} interface. The variation weights are labelled as \texttt{SoftPhysics.v1}, \texttt{SoftPhysics.v2}, 
    etc., where \texttt{v1} corresponds to the first specified variation, i.e.\ $\vec{p}_1$.


\section{Auxiliary material}\label{app2}

\begin{figure}[h]
    \centering
    \includegraphics[width=1.0\textwidth]{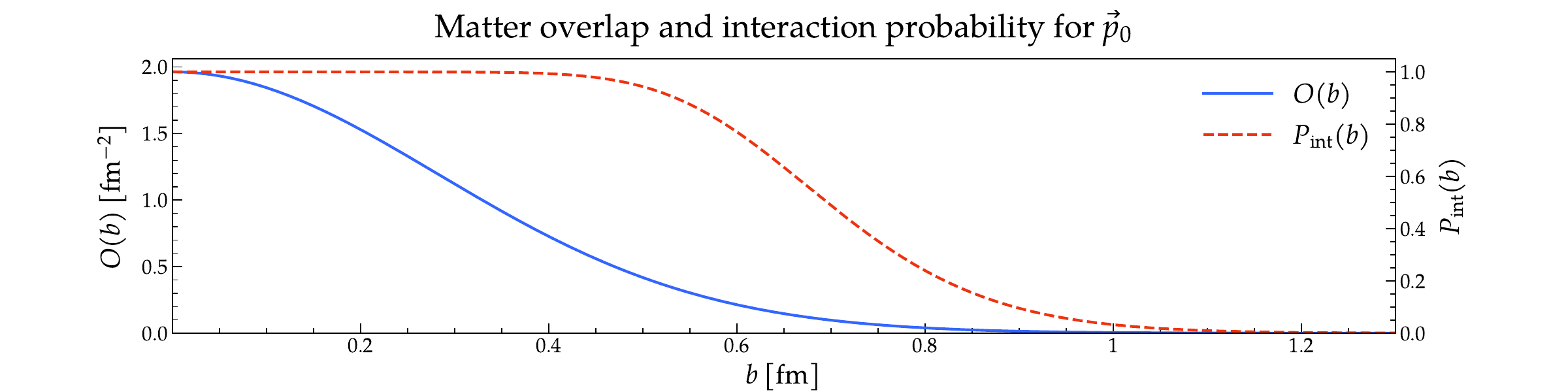}
    \caption{Overlap function $O(b)$~(\ref{eq:overlap}) and interaction probability $P_{\rm int}(b)$~(\ref{eq:pint}) for the nominal parameter set $\vec{p}_0$ (cf. Section \ref{sec:tuning_method}), as a function of the impact parameter $b$. 
    For this parameter set, the $k$ factor is fixed to $k = 0.31$.}
    \label{fig:overlap_and_pint}
\end{figure}

\begin{figure}[h]
    \centering
    \includegraphics[width=0.70\textwidth]{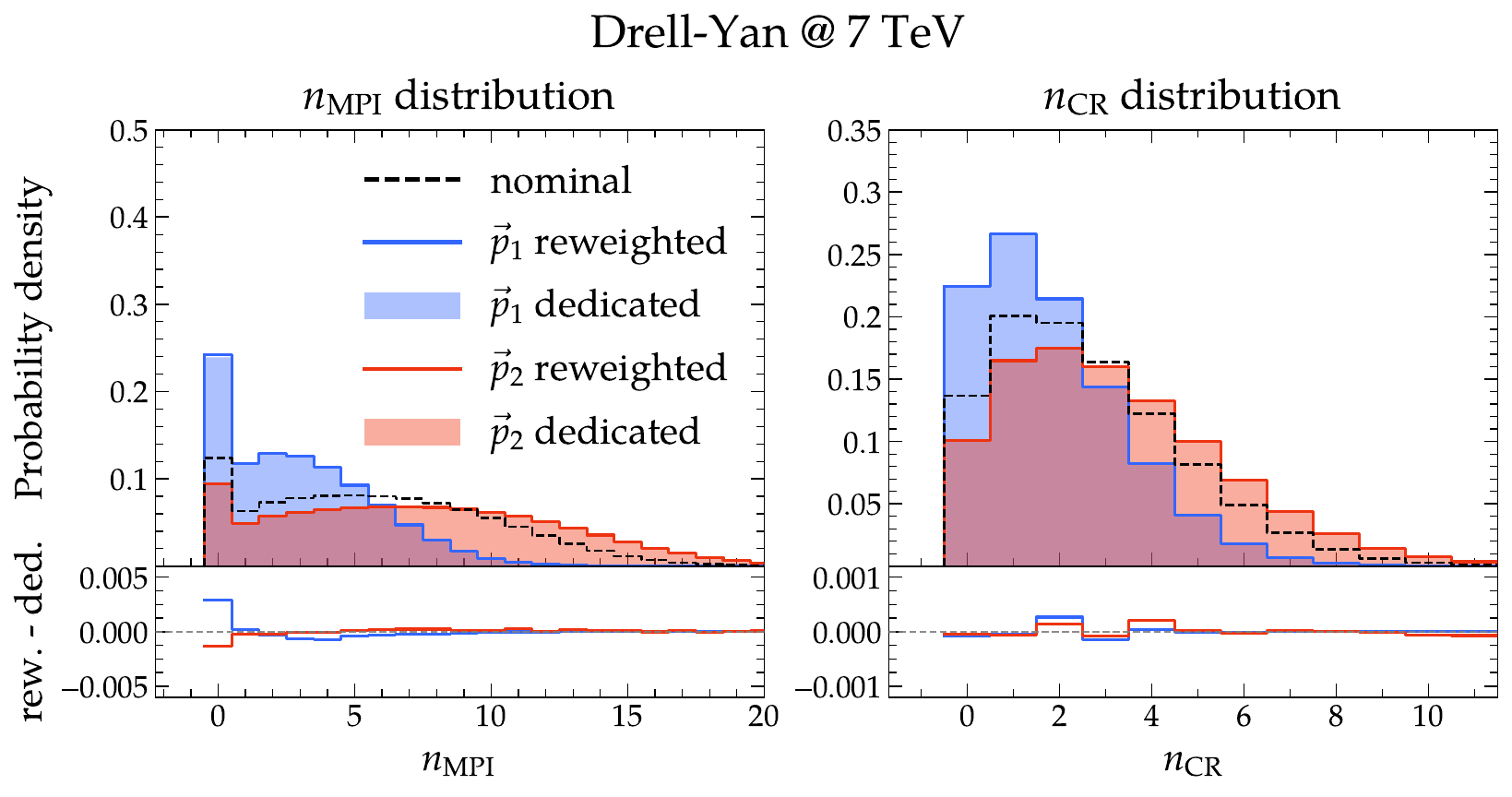}
    \caption{Distributions of 
    the number of multiple parton interactions $n_{\rm MPI}$ and the number of colour reconnections $n_{\rm CR}$ in dijet events at $\sqrt{s}=7\,\text{TeV}$
    (the impact-parameter distribution is independent for the same colliding particles and is not repeated here, cf. Figure~\ref{fig:DY_distributions}). 
    For the MPI variations (first panel), the parameter sets are 
    $\vec{p}^{\rm\:MPI}_1 = (R_1 = 0.75~\mathrm{fm},\, \alpha_1 = 0.7,\, \sigma_{\rm ND}^{\rm norm} = 0.7,\, p_{\perp,\min}^{\rm ref} = 1.3~\mathrm{GeV})$ and 
    $\vec{p}^{\rm\:MPI}_2 = (0.95~\mathrm{fm},\,0.3,\,0.3,\,1.0~\mathrm{GeV})$, 
    while for the CR variation (third panel) the parameter sets are $\vec{p}^{\rm\:CR}_1 = (\eta_Q = 0.35)$ and $\vec{p}^{\rm\:CR}_2 = (0.65)$. 
    The nominal parameter set is given by
    $\vec{p}_0 = (R_1 = 0.85~\mathrm{fm},\, \alpha_1 = 0.5,\, \sigma_{\rm ND}^{\rm norm} = 0.5,\, p_{\perp,\min}^{\rm ref} = 1.0~\mathrm{GeV},\, \eta_Q = 0.5)$. 
    Each panel shows both the dedicated and the reweighted distributions, the latter are obtained by applying the reweighting weights to the nominal sample. 
    The lower panels display the differences between both results.}
    \label{fig:Jets_distributions}
\end{figure}

\begin{figure}
    \centering
    \includegraphics[width=1.0\textwidth]{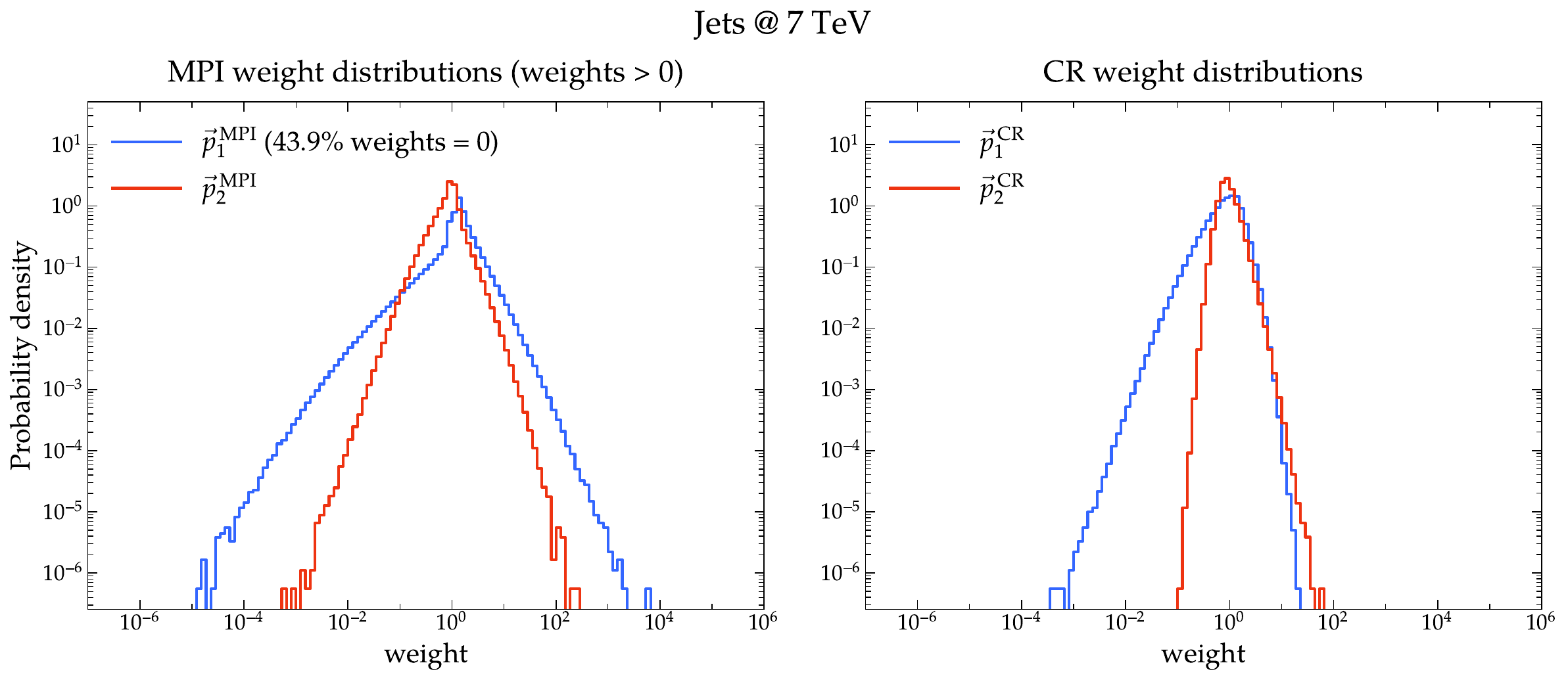}
    \caption{Distributions of the reweighted event weights obtained from the dijet process validation runs at $\sqrt{s}=7\,\text{TeV}$. 
    The two panels show the weight distributions corresponding to the MPI (left) and CR (right) parameter variations used in Fig.~\ref{fig:Jets_distributions}.}
    \label{fig:Jets_weight_distribution}
\end{figure}

\begin{figure}
\centering
\begin{minipage}[t]{0.49\textwidth}
    \centering
    \includegraphics[width=1.0\textwidth]{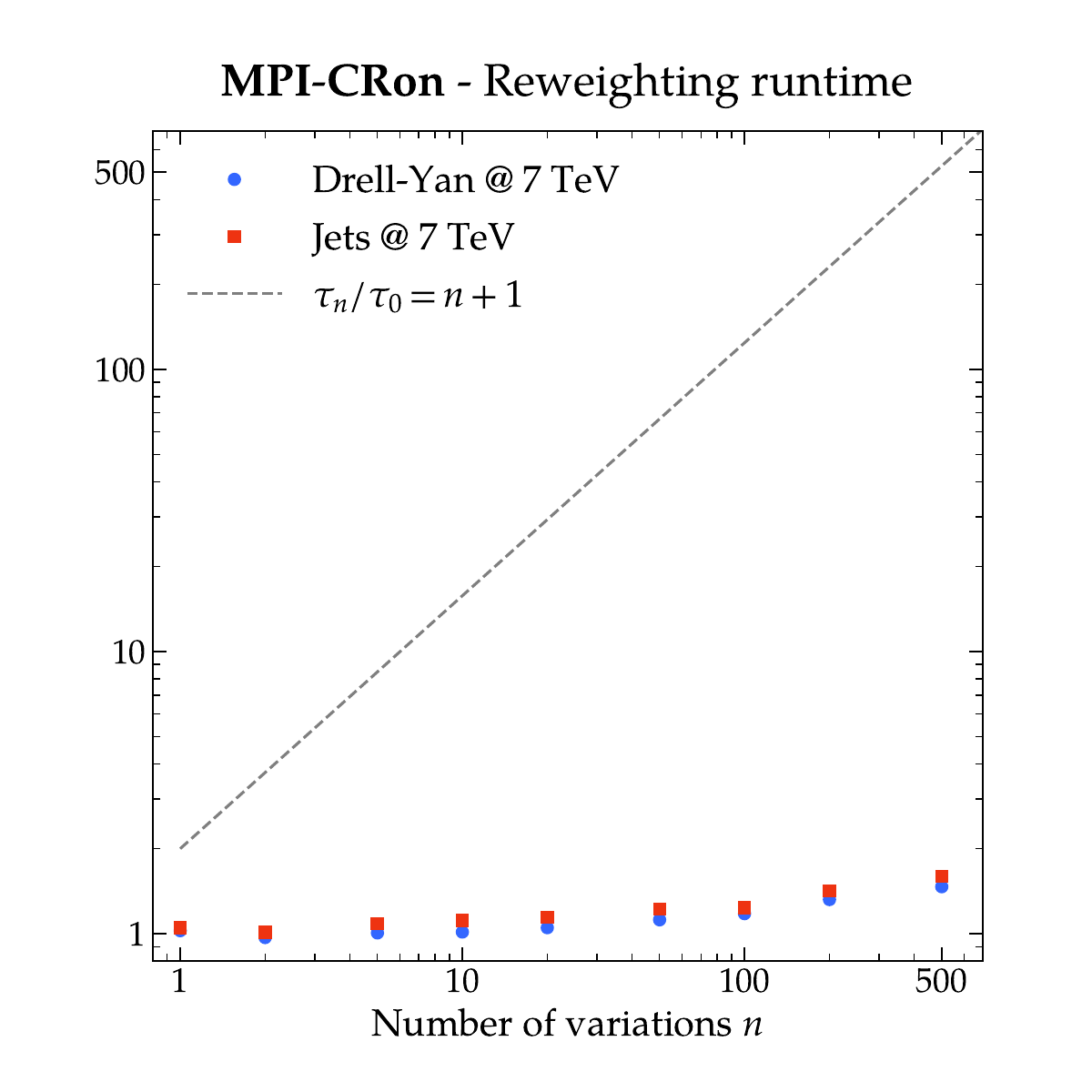}
\end{minipage}\hfill
\caption{The ratio of the CPU time required for a reweighting event with $n$ additional variations, $\tau_n$, to the time 
required for a run without variations, $\tau_0$ (excluding initialisation time). All runs use the same nominal parameter 
set $\vec{p}_0$ described in Section~\ref{sec:tuning} and the variations were 
randomly sampled from the parameter ranges given in Table~\ref{tab:paramstunerange} (\textbf{MPI-CRon}). All event simulations were performed on identical hardware and the
runtime was averaged over multiple executions. For reference, the dashed line indicates the linear scaling, $\tau_n/\tau_0 = n + 1$, i.e.\
the limit when there would be no time saving from reweighting.}
\label{fig:runtime}
\end{figure}

\begin{figure}
\centering
\begin{minipage}[t]{0.49\textwidth}
    \centering
    \includegraphics[width=1.0\textwidth]{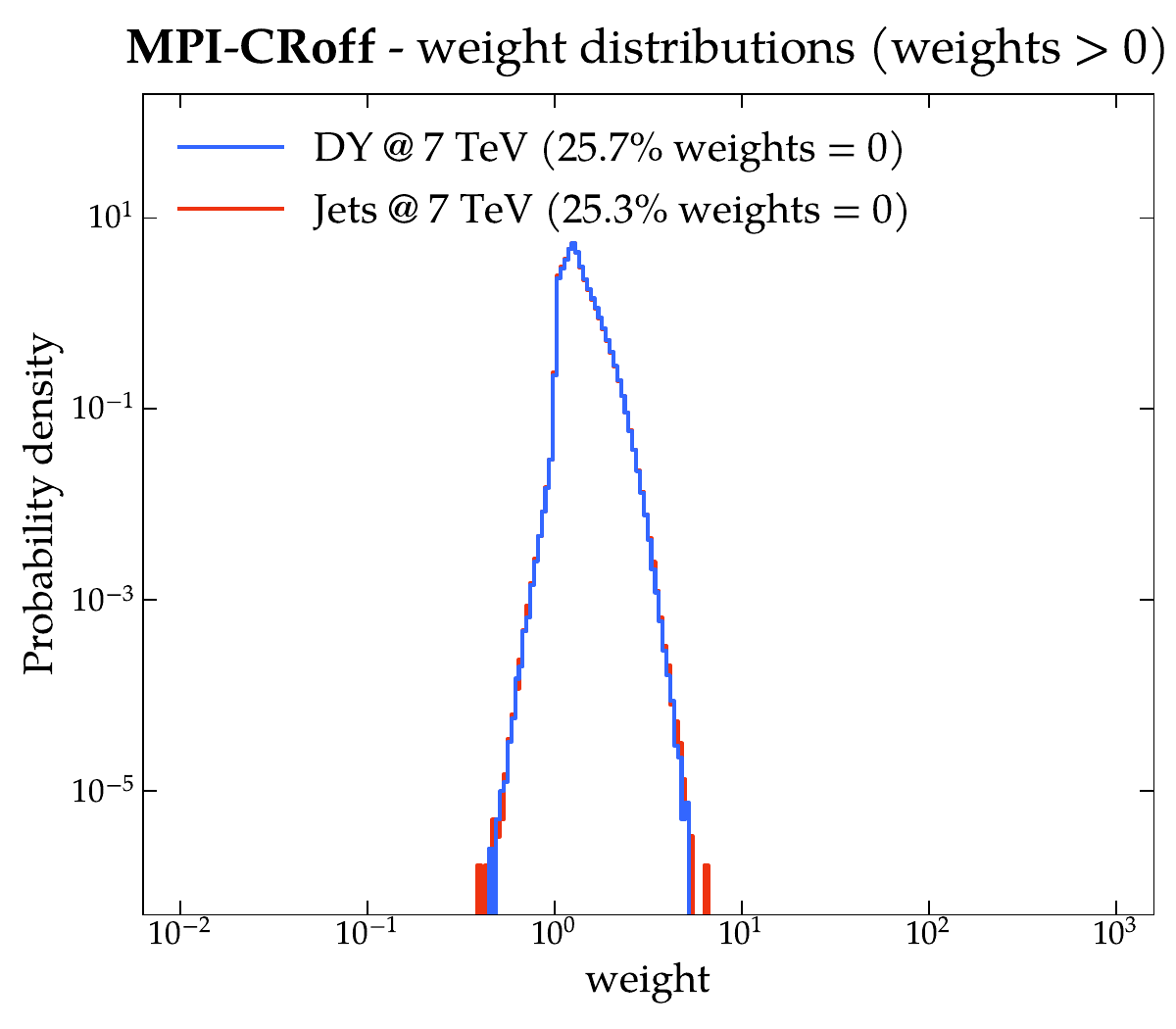}
\end{minipage}\hfill
\begin{minipage}[t]{0.49\textwidth}
    \centering
    \includegraphics[width=1.0\textwidth]{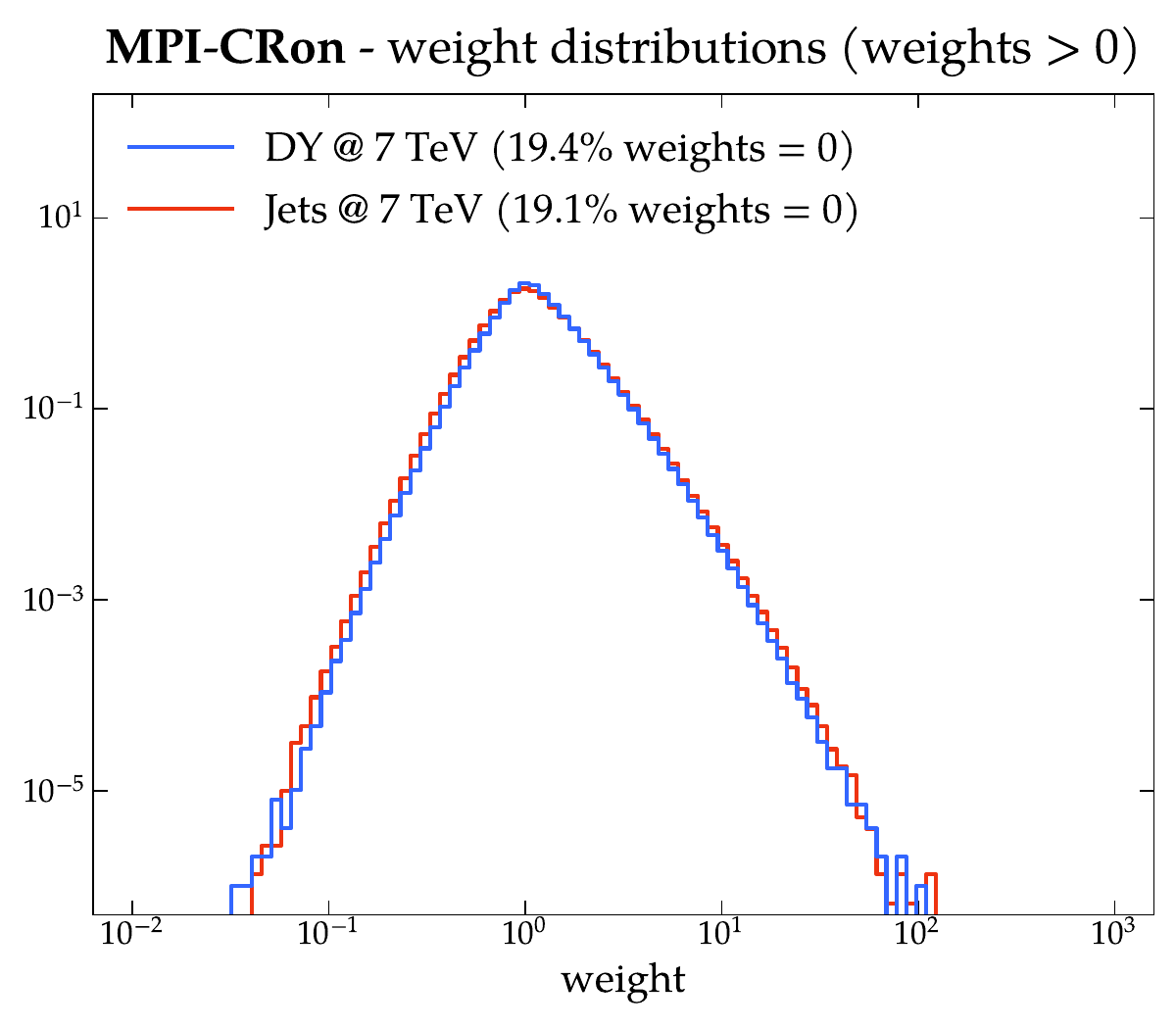}
\end{minipage}
\caption{Distribution of the total reweighting event weights obtained from the best-tune parameter runs (Figures~\ref{fig:results_DY_CR_off}-\ref{fig:results_Jets_CR_on}) for \textbf{MPI-CRoff} (left) 
  and \textbf{MPI-CRon} (right) tunes. The distributions are shown for the Drell--Yan and jet samples, respectively.
}
\label{fig:total_weight_distribution}
\end{figure}

\begin{figure}
    \centering
    \includegraphics[width=1.0\textwidth]{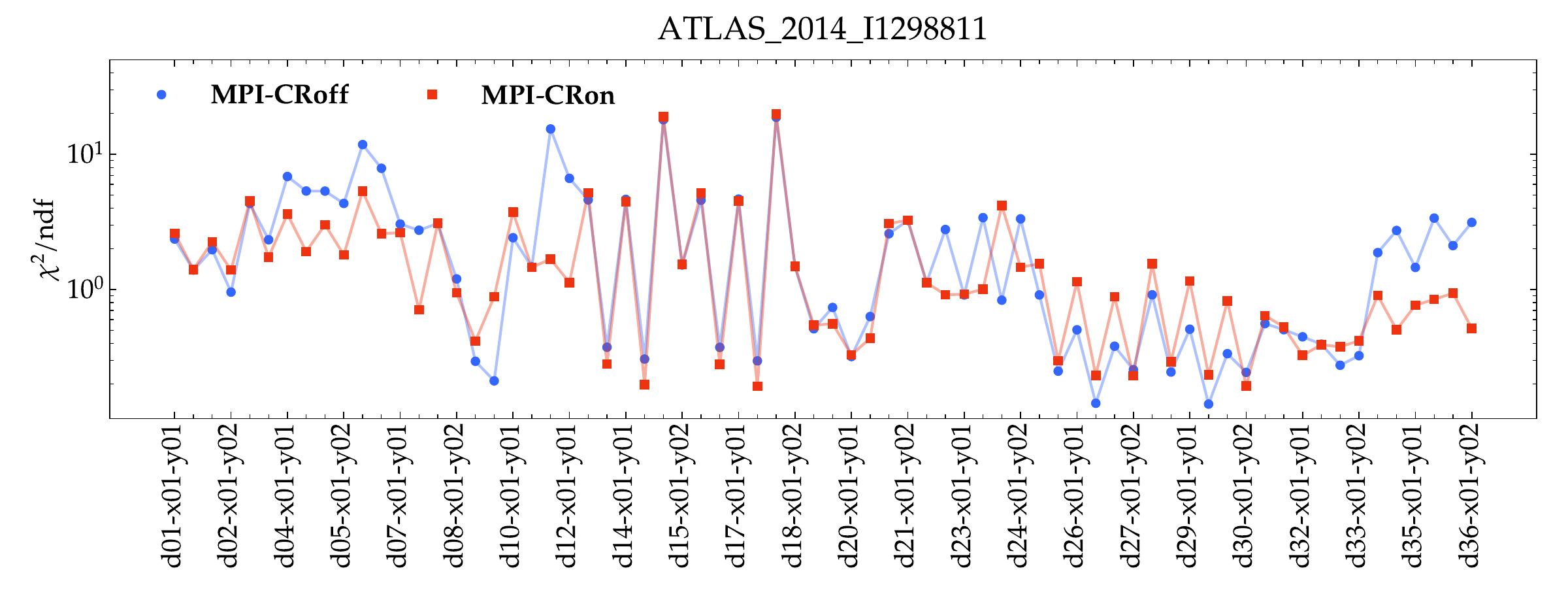}
    \includegraphics[width=1.0\textwidth]{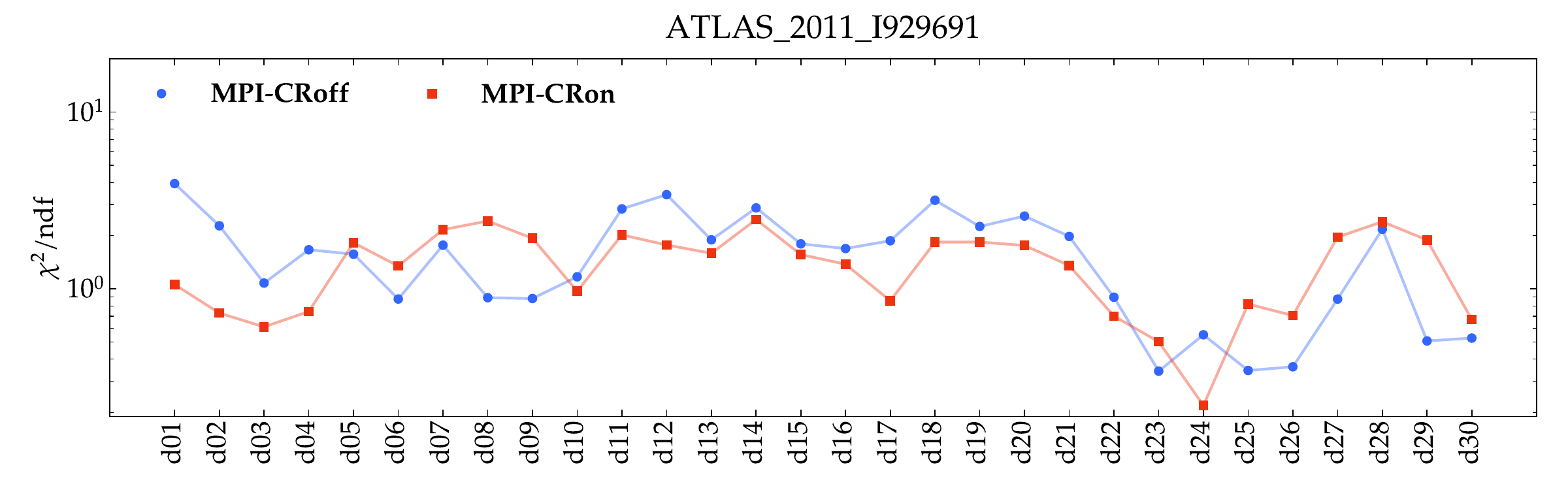}
    \includegraphics[width=1.0\textwidth]{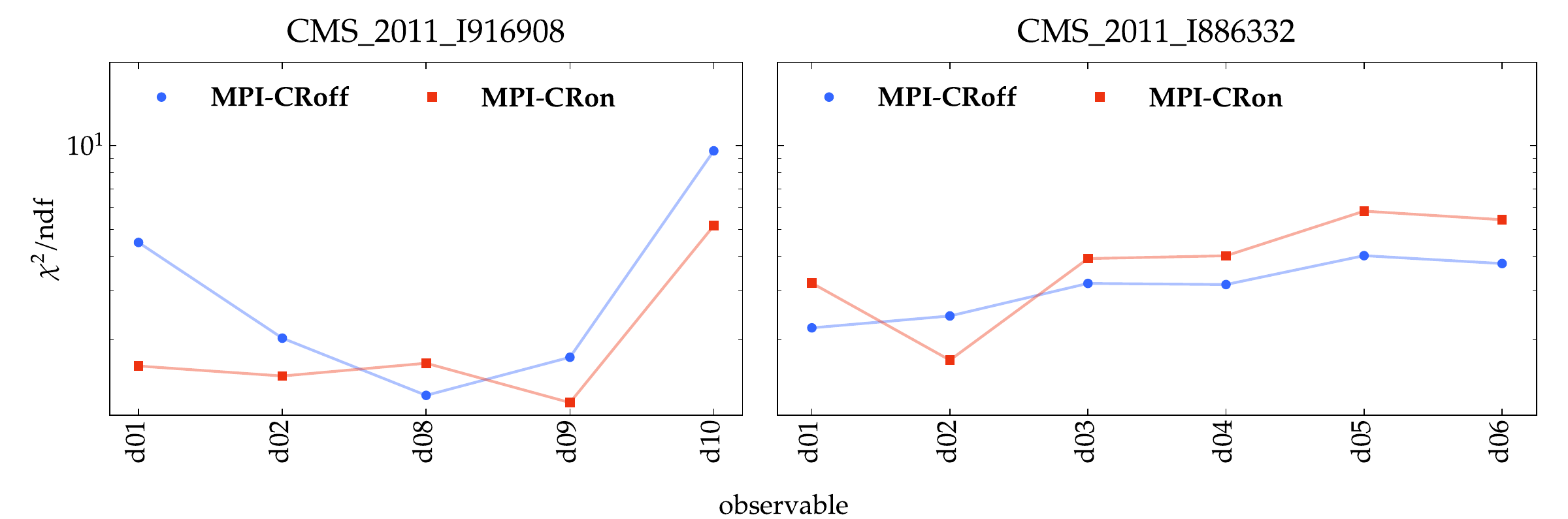}
    \caption{$\chi^2/{\rm ndf}$ values per observable for all jet analyses considered in the 
    \textbf{MPI-CRoff} and \textbf{MPI-CRon} tunes, excluding \texttt{CMS\_2011\_S9215166}, 
    which contains only a single valid observable.}
    \label{fig:chi2_plots_Jets}
\end{figure}

\bibliographystyle{amsunsrt_modp}
\bibliography{main}

\end{document}